\documentclass[sigconf,nonacm]{aamas} 

\usepackage{pbalance} % for balancing columns on the final page
\usepackage{bbm}
\usepackage{graphicx}
\usepackage{subcaption}
\usepackage{multirow} 
\usepackage{algorithm}
\usepackage{algpseudocode}

\newcommand{\algoname}{\text{MIRL}}

\setcopyright{none}
\acmDOI{}
\acmISBN{}

\title{Preference-Guided Cooperation in\\ Multi-Objective Multi-Agent Reinforcement Learning}
\author{Qiyue Xia}
\affiliation{
  \department{School of Informatics}
  \institution{University of Edinburgh}
  \city{Edinburgh}
  \country{United Kingdom}}
\email{qiyue.xia@ed.ac.uk}

\author{J.~Michael Herrmann}
\affiliation{
  \department{School of Informatics}
  \institution{University of Edinburgh}
  \city{Edinburgh}
  \country{United Kingdom}}
\email{michael.herrmann@ed.ac.uk}

\begin{abstract}
Multi-objective multi-agent reinforcement learning (MOMARL) addresses agent coordination and objective trade-offs simultaneously while optimising overall system performance.
Existing work tends to treat multi-agent cooperation and multi-objective optimisation as largely independent components of a learning problem, which amounts to an exorbitant complexity except in trivial cases.
We study here whether the structure provided by multiple objectives can itself be exploited to facilitate cooperation, rather than being treated as an additional optimisation challenge.
We formulate MOMARL at both the team and individual levels and introduce objective-wise counterfactual credit assignment to evaluate each agent’s contribution across the team objectives. 
We propose Meta-Gradient Informed Multi-Objective Multi-Agent Reinforcement Learning (MIRL), with static (MIRL-S) and dynamic (MIRL-D) variants. 
By incorporating meta-gradient information, MIRL enables agents to learn how to prioritise objectives individually while collectively optimising the system-level objective, allowing the multi-objective structure to directly guide cooperation.
Experiments on a multi-objective multiple particle environment show that our methods improve Pareto-front quality over baselines, with the dynamic variant achieving the strongest performance.

\end{abstract}


\newcommand{\BibTeX}{\rm B\kern-.05em{\sc i\kern-.025em b}\kern-.08em\TeX}

\begin{document}

%%% The following commands remove the headers in your paper. For final 
%%% papers, these will be inserted during the pagination process.

%\pagestyle{fancy}
%\fancyhead{}

%%% The next command prints the information defined in the preamble.

\sloppy

\maketitle 

% arXiv: remove conference headers on all pages
\fancypagestyle{standardpagestyle}{%
    \fancyhf{}
    \renewcommand{\headrulewidth}{0pt}
    \renewcommand{\footrulewidth}{0pt}
    \fancyfoot[C]{\thepage}
}
\pagestyle{standardpagestyle}

% Remove first-page conference header
\thispagestyle{empty}

%%%%%%%%%%%%%%%%%%%%%%%%%%%%%%%%%%%%%%%%%%%%%%%%%%%%%%%%%%%%%%%%%%%%%%%%

\section{Introduction}
Many real-world decision-making problems naturally involve both multiple decision makers and multiple objectives. For example, autonomous vehicles~\citep{ShalevShwartz2016SafeMR}, drones~\citep{hung2016q}, smart homes~\citep{tittaferrante2022multiadvisor}, and collaborative industrial systems~\citep{peng2021facmac} require several autonomous entities to interact with a shared environment while simultaneously considering different and potentially conflicting performance criteria. These requirements motivate two important extensions of conventional reinforcement learning. Multi-Agent Reinforcement Learning (MARL) addresses decision-making with multiple simultaneously interacting agents, where the learning problem is complicated by non-stationarity caused by other agents simultaneously interacting with the environment and adapting their policies, partial observability due to agents having access to only local information, and the need to develop effective coordination among agents. Multi-Objective Reinforcement Learning (MORL), in contrast, extends reinforcement learning beyond the optimisation of a single scalar objective to problems involving multiple, potentially conflicting objectives. Rather than searching for one optimal policy, MORL commonly aims to approximate a Pareto front consisting of policies that represent different trade-offs among objectives, allowing solutions to be selected according to different preferences.

Multi-Objective Multi-Agent Reinforcement Learning (MOMARL) brings these two dimensions together, providing a more comprehensive modelling framework for complex decision-making environments in which both interactions among multiple decision makers and trade-offs among multiple objectives must be considered. 
Although MARL and MORL have each developed into substantial research areas, their intersection remains comparatively under-explored. 
% MOMARL therefore represents an important but still developing direction for modelling complex cooperative decision-making systems.
Recent work has gradually extended the scope of MOMARL through dedicated deep learning methods~\cite{hu2023momix}, standardised benchmarks~\citep{felten2026momaland,geng2025mosmac}, and more flexible formulations of agent utilities and preferences~\citep{li2026achieving-moma-heterogeneity}. 
% However, these existing approaches have largely treated these two aspects as parallel dimensions of the learning problem: multi-agent mechanisms address interactions and coordination among agents, while multi-objective mechanisms primarily represent and optimise trade-offs among system-level objectives.
However, these existing approaches have largely treated the multi-objective component as an additional optimisation dimension built on top of the multi-agent problem, rather than as a source of structure that can itself inform and shape the broader process of multi-agent learning.

In this work, we take a different perspective on the MOMARL formulation and demonstrate that the multi-objective structure itself can be exploited to facilitate multi-agent cooperation.
% A multi-objective system provides richer information than a scalar team reward, as the performance of the team can be decomposed across different objectives. This raises the possibility that individual agents need not always respond to these objectives in exactly the same way. Although the overall system must ultimately satisfy a user-specified trade-off among its objectives, individual agents may occupy different roles, contribute differently to the objectives, and face changing requirements as the cooperative interaction evolves. 
% From this perspective, 
We aim to provide agents with a higher-level adaptive capability, allowing them to adjust how they contribute to the team according to the current cooperative situation.
% To realise this adaptive capability, we proposed a bi-level preference formulation. The externally specified team preference defines the desired system-level trade-off, while each agent has a learnable individual preference that determines how it responds to the different objectives during policy learning. 
To realise this capability, we propose a bi-level preference formulation
with a team preference and learnable individual preferences. 
Since these individual preferences influence how the policy itself is updated, learning them naturally forms a learning-to-learn problem. We therefore derive a novel meta-gradient mechanism that adapts the individual preferences according to how the resulting policy updates affect subsequent team-level performance.

Building on this framework, we introduce Meta-Gradient Informed Multi-Objective Multi-Agent Reinforcement Learning (\algoname{}), a novel algorithm that enables agents to adapt their learning across multiple objectives while remaining aligned with the system-level objective. 
Our main contributions are as follows:

(1) We formulate cooperative MOMARL from both team and individual levels, where a team preference defines the desired system-level objective trade-off and learnable individual preferences provide adaptive guidance for individual behaviour. 

(2) We introduce objective-wise counterfactual credit assignment, which evaluates each agent's contribution separately across the team objectives and provides objective-specific learning signals.

(3) We derive a novel meta-gradient preference adaptation formulation, treating individual preferences as meta-variables that shape policy learning and optimising them according to their effect on subsequent team-level performance. We develop a static variant, \algoname{}-S, and a dynamic variant, \algoname{}-D, which further adapts preferences according to the current cooperative context.

(4) We conduct comprehensive experiments in Multi-Agent Particle Environment. The results demonstrate that our methods consistently improve Pareto-front quality over a range of baselines, with the dynamic variant achieving the strongest overall performance.

\section{Background}
\subsection{Cooperative Multi-Agent Reinforcement Learning}

Multi-Agent Reinforcement Learning (MARL) extends the traditional reinforcement learning framework to problems in which multiple agents simultaneously interact with a shared environment. In this work, we focus on cooperative and partially observable settings, where agents receive local observations and share a common team reward. Such a setting can be formalised as a Decentralised Partially Observable Markov Decision Process (Dec-POMDP).

\begin{definition}{Decentralised Partially Observable Markov Decision Process.} 
A Dec-POMDP is defined by the tuple 
\((\mathcal{N},\mathcal{S},\{\mathcal{O}^i\}_{i=1}^{N},\allowbreak \{\mathcal{A}^i\}_{i=1}^{N},P,R,\mathcal{Z},\gamma)\), 
where
$\mathcal{N}=\{1,\ldots,N\}$ is the set of agents,
$\mathcal{S}$ is the global state space,
$\mathcal{O}^i$ and $\mathcal{A}^i$ are the observation and action spaces of agent $i$, respectively,
$\mathcal{O}=\mathcal{O}^1\times\cdots\times\mathcal{O}^N$ and
$\mathcal{A}=\mathcal{A}^1\times\cdots\times\mathcal{A}^N$ denote the joint observation and joint action spaces,
$P(s'|s,\mathbf{a})$ is the state transition probability under joint action $\mathbf{a}$,
$R:\mathcal{S}\times\mathcal{A}\rightarrow\mathbb{R}$ is the shared team reward function,
$\mathcal{Z}(\mathbf{o}|s',\mathbf{a})$ is the observation function specifying the probability of receiving joint observation $\mathbf{o}\in\mathcal{O}$ after joint action $\mathbf{a}$ results in state $s'$, and $\gamma$ is the discount factor.
\end{definition}

At each timestep $t$, agent $i$ receives a local observation $o_t^i\in\mathcal{O}^i$ and selects an action $a_t^i\sim\pi^i(\cdot\mid o_t^i)$. The individual actions form the joint action $\mathbf{a}_t=(a_t^1,\ldots,a_t^N)$, which determines the subsequent state and the shared team reward, while the agents receive their corresponding local observations through the observation function $\mathcal{Z}$. We denote the collection of individual policies as the joint policy $\boldsymbol{\pi}=(\pi^1,\ldots,\pi^N)$.

Under the cooperative setting considered here, all agents share the same team reward. The performance of a joint policy $\boldsymbol{\pi}$ can be represented by its expected discounted team return,
\[
V^{\boldsymbol{\pi}}(s)
=
\mathbb{E}_{\boldsymbol{\pi}}
\left[
\sum_{t=0}^{\infty}\gamma^t R(s_t,\mathbf{a}_t)\mid s_0=s
\right].
\]
The learning problem is therefore to obtain individual policies whose joint behaviour achieves high shared team performance.

Unlike single-agent reinforcement learning, each agent influences both the environment and the learning process experienced by the other agents. This introduces challenges such as non-stationarity caused by simultaneously adapting policies, partial observability due to locally available information, and the need for agents to learn coordinated behaviours.

\subsection{Multi-Objective Multi-Agent Reinforcement Learning}

Multi-Objective Multi-Agent Reinforcement Learning (MOMARL) extends MARL to problems involving multiple objectives. In this work, we consider the cooperative team-reward setting, where all agents share the same vector-valued reward and collectively contribute to the performance of the multi-agent system.
Following the standard cooperative formulation, this setting can be represented as a multi-objective decentralised partially observable Markov decision process (MO-Dec-POMDP).

\begin{definition}{Multi-Objective Decentralised Partially Observable Markov Decision Process.}
A cooperative MO-Dec-POMDP with $N$ agents is defined by the tuple $(\mathcal{N},\mathcal{S},\{\mathcal{O}^i\}_{i=1}^{N},\{\mathcal{A}^i\}_{i=1}^{N},P,R,\mathcal{Z},\gamma)$, 
where $\mathcal{N}=\{1,\ldots,N\}$ is the set of agents,
$\mathcal{S}$ is the global state space,
$\mathcal{O}^i$ and $\mathcal{A}^i$ are the observation and action spaces of agent $i$, respectively,
$\mathcal{O}=\mathcal{O}^1\times\cdots\times\mathcal{O}^N$ and $\mathcal{A}=\mathcal{A}^1\times\cdots\times\mathcal{A}^N$ denote the joint observation and joint action spaces,
$P(s'|s,\mathbf{a})$ is the state transition probability under joint action $\mathbf{a}$,
$R:\mathcal{S}\times\mathcal{A}\rightarrow\mathbb{R}^{d}$ is the shared vector-valued team reward function with $d$ objectives,
$\mathcal{Z}(\mathbf{o}|s',\mathbf{a})$ is the observation function specifying the probability of receiving joint observation $\mathbf{o}\in\mathcal{O}$ after joint action $\mathbf{a}$ results in state $s'$, and $\gamma$ is the discount factor.
\end{definition}

At each timestep $t$, the agents select a joint action $\mathbf{a}_t$ and receive a reward vector
\(
\mathbf{r}_t
=
R(s_t,\mathbf{a}_t)
=
\bigl(R_1(s_t,\mathbf{a}_t),\ldots,R_d(s_t,\mathbf{a}_t)\bigr).
\)
The corresponding discounted team return vector is
\(
\mathbf{G}_t
=
\sum_{k=0}^{\infty}\gamma^k\mathbf{r}_{t+k},
\)
and the expected vector-valued performance of a joint policy $\boldsymbol{\pi}$ is
\[
\mathbf{V}^{\boldsymbol{\pi}}(s)
=
\mathbb{E}_{\boldsymbol{\pi}}
\left[
\mathbf{G}_t\mid s_t=s
\right].
\]

Similar to the single-agent MORL setting, MOMARL uses a preference vector
$\boldsymbol{\omega}\in\Omega$ to specify the desired trade-off among the $d$
team objectives. However, in the multi-agent setting, this preference evaluates
the collective performance produced by the joint policy
$\boldsymbol{\pi}=(\pi^1,\ldots,\pi^N)$. Under linear scalarisation, the
preference-weighted value of a joint policy is defined as
\[
V_{\boldsymbol{\omega}}^{\boldsymbol{\pi}}(s)
=
f_{\boldsymbol{\omega}}
\left(\mathbf{V}^{\boldsymbol{\pi}}(s)\right)
=
\boldsymbol{\omega}^\top\mathbf{V}^{\boldsymbol{\pi}}(s),
\]
where $\sum_{j=1}^{d}\omega_j=1$ and $\omega_j\geq0$. For a given preference
$\boldsymbol{\omega}$, the corresponding MOMARL problem therefore seeks a joint
policy that maximises the scalarised team value,
\[
\boldsymbol{\pi}_{\boldsymbol{\omega}}^{*}
=
\arg\max_{\boldsymbol{\pi}}
V_{\boldsymbol{\omega}}^{\boldsymbol{\pi}}(s).
\]
Different preference vectors may consequently lead to different optimal joint
policies and different trade-offs among the team objectives.

\subsection{Pareto Optimality in MOMARL}

% In multi-objective optimisation, the concept of optimality differs from the single-objective case. Typically, no single policy simultaneously maximises all objectives due to inherent trade-offs. Without any additional information about the user’s preference, there can now be multiple possibly optimal solutions. In the following, we introduce several useful definitions for possibly optimal policies.
% The concept of Pareto optimality introduced in Section~1.2.2 extends naturally to the multi-agent setting. While a solution in MORL corresponds to a policy $\pi$, a solution in MOMARL corresponds to a joint policy $\boldsymbol{\pi}$ whose performance is determined by the collective behaviour of all agents.

In multi-objective optimisation, the concept of optimality differs from the single-objective case. Typically, no single solution simultaneously maximises all objectives due to inherent trade-offs. Without any additional information about the user’s preference, there may therefore be multiple possibly optimal solutions. 
Thus, in MORL, Pareto optimality is introduced as a solution concept for possibly optimal policies. This concept extends naturally to the multi-objective multi-agent setting. 

\begin{definition}{Pareto optimality in MOMARL.}
A joint policy $\boldsymbol{\pi}$ is said to \emph{dominate} another joint policy $\boldsymbol{\pi}'$ if and only if $\forall j\in\{1,\ldots,d\},\,V_j^{\boldsymbol{\pi}}(s)\geq V_j^{\boldsymbol{\pi}'}(s)$ and $\exists k\in\{1,\dots,d\},\,V_k^{\boldsymbol{\pi}}(s)>V_k^{\boldsymbol{\pi}'}(s)$, where $V_j^{\boldsymbol{\pi}}(s)$ denotes the expected discounted team return for objective $j$ under joint policy $\boldsymbol{\pi}$. A joint policy $\boldsymbol{\pi}^{*}$ is \emph{Pareto optimal} if and only if it is not dominated by another joint policy. The set of all Pareto optimal joint policies forms the \emph{Pareto set}, $\Pi_P=\{\boldsymbol{\pi}\mid\boldsymbol{\pi}\text{ is Pareto optimal}\}$. The corresponding set of expected vector returns is termed the \emph{Pareto front}, $\mathcal{P}=\{\mathbf{V}^{\boldsymbol{\pi}}(s)\mid\boldsymbol{\pi}\in\Pi_P\}$.
\end{definition}

Since obtaining the true Pareto set is generally intractable in complex MOMARL problems, the practical objective is to construct a finite set of joint policies that closely approximates the Pareto front, allowing a suitable joint policy to be selected according to the desired trade-off among objectives.

\section{Related Work}
In recent years, deep Reinforcement Learning has shown great promise in solving complex decision-making problems, enabling significant advancements across a wide range of applications~\citep{silver2016mastering, levine2016end}.
In real-world scenarios, however, problems often feature multiple, potentially conflicting objectives and involve multiple agents interacting in a shared environment, which raises challenging problems in the RL domain.
A growing number of studies have investigated these challenges from different perspectives. We introduce relevant works in the following.

MORL extends RL methods to handle multiple, potentially conflicting objectives, and existing approaches can be broadly grouped into single-policy methods and multi-policy methods for approximating the Pareto front.
Single-policy approaches typically convert the multi-objective problem into a single-objective task using a predefined preference or weighting scheme, including linear scalarization \cite{barrett2008learning,vamplew2011empirical}, and more advanced scalarization functions such as Chebyshev methods~\citep{van2013scalarized}, hypervolume-based approaches~\citep{zhang2020random} and the addition of concave terms to rewards~\citep{lu2023multi}. 
Multi-policy MORL strategies approximate the Pareto front by learning a diverse collection of policies. One direction is developing single models conditioned on preferences~\citep{abels2019dynamic, kyriakis2022pareto}. Further techniques for generating policy sets involve Generalised Policy Improvement (GPI) for sample-efficient learning~\citep{alegre2023sample} or the development of transferable policy components using representations like successor features~\citep{alegre2022optimistic}. 
Besides these approaches, several studies have explored the structural relationships among Pareto-optimal solutions from other perspectives, such as navigation along the Pareto set~\citep{ye2022pareto}, exploitation of regularities in policy parameter space~\citep{xu2020prediction}, and policy manifolds~\citep{parisi2016multi,xia2025interpretability} to improve the efficiency of Pareto-front reconstruction. 

MARL considers learning problems involving multiple interacting agents. In this paper, we focus on the domain of cooperative MARL, where several agents must cooperate to accomplish a shared task. 
Independent learning is one of the simplest MARL paradigms, in which each agent receives local observations and applies a reinforcement learning algorithm independently~\citep{tan1993multi}. Although such approaches can be effective in simple settings~\citep{claus1998dynamics,gupta2017cooperative}, they suffer from non-stationarity because other agents are simultaneously interacting with the environment and adapting their policies~\citep{hernandez2019survey}. 
The centralised training with decentralised execution (CTDE) framework has therefore widely adopted to exploit additional global information during training while retaining decentralised policies at execution time, as in MADDPG~\citep{lowe2017multi}, MAAC~\citep{iqbal2019actor}, and MAPPO~\citep{yu2022MAPPO}. 
Credit assignment is another major challenge in cooperative MARL. When agents share a team reward, it can be difficult to determine the contribution of each individual agent to the overall outcome, potentially leading to the \textit{lazy agent} problem, where poorly contributing agents may still receive high rewards due to the successful behaviour of their teammates. To address this issue, value-decomposition methods such as VDN~\citep{sunehag2018value} and QMIX~\citep{tabish2020qmix} decompose the joint value function into agent-level utilities, while COMA~\citep{foerster2018counterfactual} uses a counterfactual baseline to evaluate each agent's contribution to the team outcome. 
Cooperative MARL has also explored mechanisms for facilitating coordinated behaviours, including learned communication~\citep{foerster2016learning,sukhbaatar2016learning} and emergent role specialisation~\citep{wang2020roma}.

MOMARL studies multi-agent decision-making problems with multiple, potentially conflicting objectives, where interactions among agents and trade-offs among objectives must be considered jointly.
Early studies mainly addressed application-specific problems, often reducing multiple objectives to a scalar learning objective, as in cooperative traffic signal control~\citep{khamis2014adaptive}, or considering independent agent learning, as in smart-home energy control~\citep{tittaferrante2022multiadvisor}. 
Later work moved beyond application-specific formulations towards more general learning mechanisms and conceptual analyses for MOMARL.
Potential-based reward shaping and difference rewards were investigated to address credit assignment in cooperative multi-objective settings~\citep{mannion2018reward}. 
A systematic utility-based analysis further characterised multi-objective multi-agent decision-making problems according to their reward structures, utility functions, and corresponding solution concepts~\citep{ruadulescu2020multi}. 
More recently, dedicated deep MOMARL methods have been developed. MO-MIX~\citep{hu2023momix} extends value-decomposition methods to the multi-objective setting, using preference-conditioned value functions to learn cooperative policies for different objective trade-offs. 
The development of standardised benchmarks has also supported broader evaluation of MOMARL methods, with MOMAland~\citep{felten2026momaland} providing a diverse benchmark suite and MOSMAC~\citep{geng2025mosmac} introducing sequential multi-objective tasks with longer temporal horizons.
Recent work has further considered continuous state and action spaces through MOMA-AC~\citep{callaghan2025momaac}, and heterogeneous utility functions across agents~\citep{li2026achieving-moma-heterogeneity}.

The above works have explored the MOMARL problem from different perspectives, but have mainly considered multiple objectives from the perspective of system-level optimisation. Comparatively less attention has been paid to how multiple objectives can shape the learning and cooperation of individual agents. 
We move beyond a solely system-level view by incorporating an individual-level learning-to-learn perspective into MOMARL, providing agents with the adaptive capability to adjust how they contribute to the cooperative task. 
% From this perspective, rather than viewing the multi-objective component only as an additional optimisation difficulty, we exploit the structure provided by multiple objectives itself to facilitate multi-agent cooperation. 
From this perspective, rather than viewing the multi-objective component only as an additional optimisation difficulty, we claim that the structure provided by multiple objectives can itself be exploited to facilitate multi-agent cooperation.

% \cite{ruadulescu2020multi}
% Foundational taxonomy / conceptual framing. 
% It analyses multi-objective multi-agent problems through reward structures and utility functions, including whether utilities are shared or individual and which solution concepts make sense. 

% \cite{hu2023momix}
% One of the key dedicated deep cooperative MOMARL algorithms.

% \cite{felten2024momaland}
% Standardised MOMARL benchmark suite
% \cite{geng2025mosmac}
% Benchmark showing difficulty of sequential/long-horizon MOMARL

% \cite{callaghan2025momaac}
% Continuous state/action MOMARL with CTDE actor–critic

% \cite{li2026achieving-moma-heterogeneity}
% Heterogeneous utilities/preferences across agents

\section{Methods}
\begin{figure}[t]
    \centering
    \includegraphics[width=0.9\linewidth]{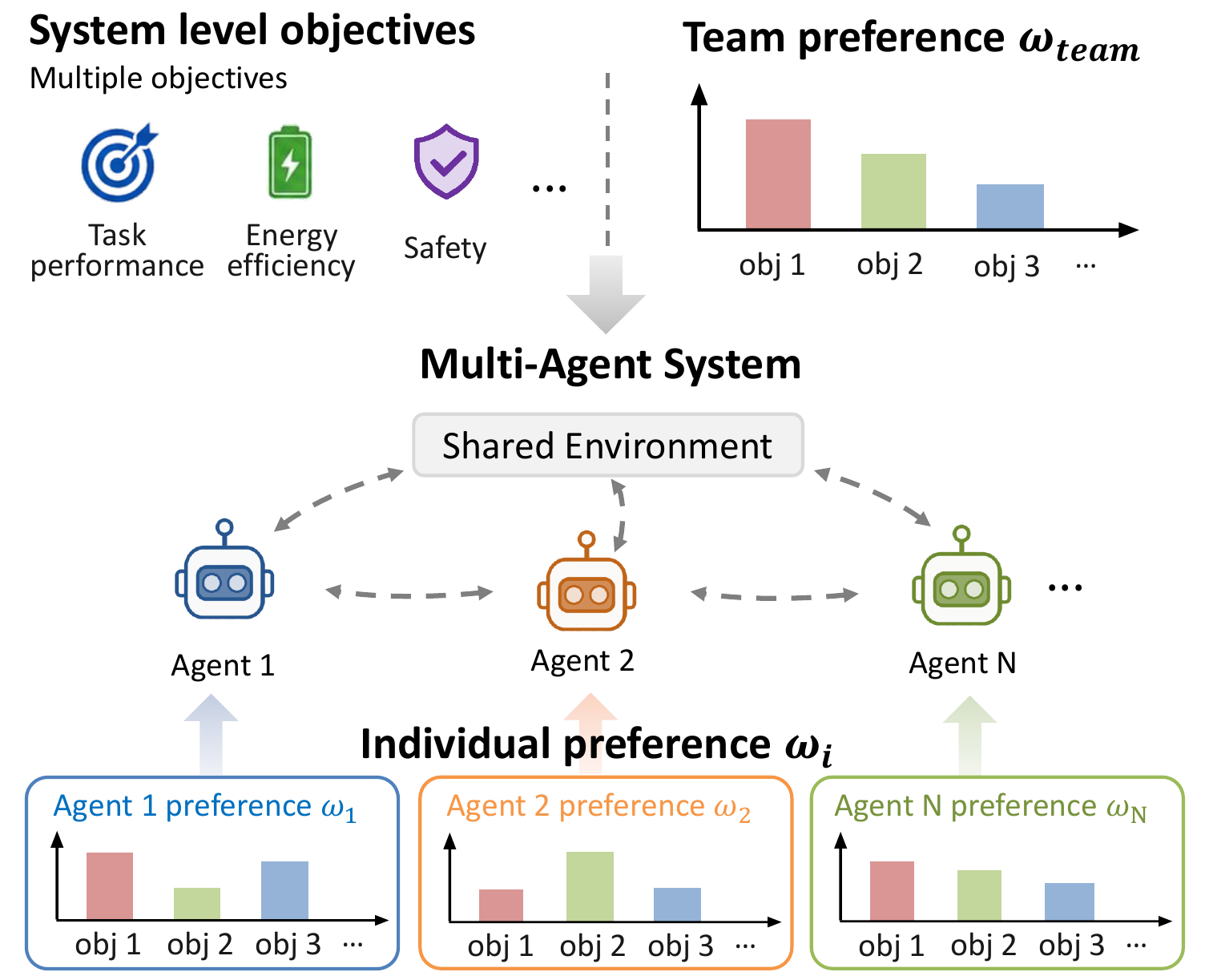}
    \caption{Team-level and individual-level preference modelling in MOMARL. 
    An externally specified team preference defines the desired system-level trade-off, while internally developed individual preferences give agents flexibility in how they contribute to the cooperative task.
    }
    \label{fig:momarl_preference}
    \Description{This figure illustrates the main features of the proposed approach including a symbolic visualisation of a possible set of objectives and the representation of the agent-related preferences in the full MOMARL setting.}
    \vspace{-0.2em}
\end{figure}

\subsection{Team-Level and Individual-Level Modelling}\label{subsec:MOMARL_modelling}

% In this work, we consider the possibility that the structure provided by multiple objectives can itself be exploited to facilitate multi-agent cooperation. Specifically, we aim to provide individual agents with a higher-level adaptive capability, allowing them to adjust how they contribute to the team according to the current cooperative situation, while the overall system remains governed by a team-level requirement.

We first model the cooperative MOMARL problem from two interconnected levels: the team level and the individual-agent level. 
At the team level, the multi-agent system is required to achieve a desired trade-off among the system objectives. We represent this externally specified global requirement using the \emph{team preference} $\boldsymbol{\omega}_{\mathrm{team}}$. 
Under linear scalarisation, for a joint policy $\boldsymbol{\pi}=(\pi^1,\ldots,\pi^N)$, the overall system performance is evaluated by
\(
V_{\boldsymbol{\omega}_{\mathrm{team}}}^{\boldsymbol{\pi}}(s)
=
\boldsymbol{\omega}_{\mathrm{team}}^\top
\mathbf{V}^{\boldsymbol{\pi}}(s).
\)

Although the system is evaluated at the team level, its behaviour is produced by multiple individual agents. 
Different agents may contribute to the system objectives in different ways depending on their local situations and interactions with other agents, and these contributions may change as the cooperative process evolves.
The team preference specifies the desired overall trade-off, but does not directly determine how each individual agent should contribute towards achieving it.
This motivates us to additionally consider the MOMARL problem from an individual-agent level. 
For each agent $i$, we introduce an \emph{individual preference}
\(
\boldsymbol{\omega}_i,
\)
which represents the agent's internal emphasis over the same set of system objectives. Unlike $\boldsymbol{\omega}_{\mathrm{team}}$, the individual preference is not an additional user-specified requirement and does not change how the overall system is evaluated. Instead, it provides an internal degree of flexibility in how an individual agent responds to the different objectives when contributing to the team.

The resulting two-level preference formulation is illustrated in Figure~\ref{fig:momarl_preference}. 
An externally specified team preference is given to the multi-agent system, while the individual preferences are learnable at the agent level, allowing the relationships among individual preferences to emerge from the learning process rather than being predefined.
To make use of this flexibility, each agent must first be able to assess its own situation within the team, in particular, to determine how much its behaviour actually contributes to the team performance. We therefore introduce objective-wise counterfactual credit assignment in the following subsection.

\subsection{Objective-Wise Counterfactual Credit Assignment}\label{subsec:counterfactual}
In cooperative MARL, a shared team reward alone does not directly reveal how much each agent's action contributes to the resulting team performance. Counterfactual Multi-Agent Policy Gradients (COMA)~\citep{foerster2018counterfactual} addresses this problem by comparing the value of the actual joint action with a counterfactual baseline in which one agent's action is varied while the actions of the other agents are kept fixed.
Within the MOMARL structure formulated above, an agent needs to assess not only its overall contribution to the team, but also how it contributes to each of the team objectives. We therefore extend counterfactual credit assignment to preserve this objective-wise structure, representing the contribution of each agent as a vector rather than immediately reducing it to a single scalar quantity. 

To obtain such objective-wise credit while retaining decentralised policies, we adopt a centralised training with decentralised execution (CTDE) structure. 
For an $ N$- agent team, during training, a centralised critic has access to the joint observation $\mathbf{o}=(o_1,\ldots,o_N)$ and joint action $\mathbf{a}=(a_1,\ldots,a_N)$, allowing the learning process to evaluate the joint behaviour of the agents. During execution, the critic is removed and each agent selects its action independently using only its locally available information.
In our implementation, the decentralised actors share a common set of parameters $\boldsymbol{\theta}$ and are distinguished by an agent identifier $e_i$. We denote the resulting policy for agent $i$ as $\pi_{\boldsymbol{\theta}}(a_{i,t}\mid h_{i,t},e_i)$, where $h_{i,t}$ represents the local information available to the agent through its recurrent policy representation. 
% Parameter sharing does not alter the decentralised execution structure, since each agent evaluates the policy independently from its own local information and identifier.

Instead of scalarising the multi-objective return before credit assignment, the centralised critic directly estimates a vector-valued action value,
\[
\mathbf{Q}_{\boldsymbol{\psi}}(\mathbf{o},\mathbf{a})
=
\left[
Q_{\boldsymbol{\psi}}^{(1)}(\mathbf{o},\mathbf{a}),
\ldots,
Q_{\boldsymbol{\psi}}^{(d)}(\mathbf{o},\mathbf{a})
\right]
\in \mathbb{R}^{d}.
\]
where each component corresponds to one team objective.

To isolate the contribution of agent $i$, we consider how the team would have performed if agent $i$ had selected a different action while the actions of all other agents remained unchanged. Let $\mathbf{a}_{-i}$ denote the actions of all agents except agent $i$. By fixing $\mathbf{a}_{-i}$ and averaging over alternative actions that agent $i$ could select according to its current policy, we obtain an agent-specific reference for its own action choice. 
We define objective-wise counterfactual baseline as
\[
\mathbf{b}_{i}^{\mathrm{cf}}
=
\mathbb{E}_{a_i' \sim \pi_{\theta_i}(\cdot\mid o_i)}
\left[
\mathbf{Q}_{\boldsymbol{\psi}}
\bigl(\mathbf{o},(a_i',\mathbf{a}_{-i})\bigr)
\right]
\in\mathbb{R}^{d}.
\]
The baseline therefore represents the expected team performance across the objectives under the same behaviour of the other agents, but with agent $i$ acting according to its current policy.

In discrete action spaces, the counterfactual expectation can be computed by summing over the possible actions of agent $i$. For continuous action spaces, however, the action space cannot be enumerated directly. We therefore approximate the counterfactual baseline using Monte Carlo sampling from the current policy. Given $K$ alternative actions
$a_i^{(k)} \sim \pi_{\theta_i}(\cdot\mid o_i), k=1,\ldots,K$,
the counterfactual baseline is approximated by
\[
\widehat{\mathbf{b}}_{i}^{\mathrm{cf}}
=
\frac{1}{K}
\sum_{k=1}^{K}
\mathbf{Q}_{\boldsymbol{\psi}}
\bigl(
\mathbf{o},
(a_i^{(k)},\mathbf{a}_{-i})
\bigr)
\in\mathbb{R}^{d}.
\]
For each sample, only the action of agent $i$ is replaced, while the actions of the other agents remain fixed. 

The contribution of the action actually taken by agent $i$ can then be measured relative to this baseline. We define the objective-wise counterfactual advantage as
\[
\mathbf{A}_{i}^{\mathrm{cf}}
=
\mathbf{Q}_{\boldsymbol{\psi}}(\mathbf{o},\mathbf{a})
-
\mathbf{b}_{i}^{\mathrm{cf}}
\in\mathbb{R}^{d},
\]
where $\widehat{\mathbf{b}}_{i}^{\mathrm{cf}}$ is used in place of $\mathbf{b}_{i}^{\mathrm{cf}}$ for continuous action spaces.

\begin{figure}[t]
    \centering
    \includegraphics[width=0.495\textwidth]{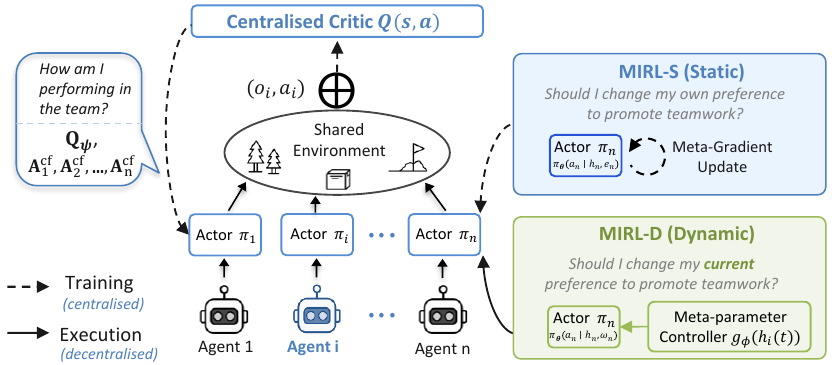}
    \caption{System overview for \algoname{}-S and \algoname{}-D.}
    \label{fig:system_overview}
    \Description{This figure highlights the differences of \algoname{}-S and \algoname{}-D as also described in Sect.~\ref{subsec:static_meta} and Sect.~\ref{subsec:dynamic_meta}.
    }
\end{figure}

Each component of $\mathbf{A}_{i}^{\mathrm{cf}}$ measures whether the action actually taken by agent $i$ improves or reduces the team performance for the corresponding objective relative to its own counterfactual baseline.
% This provides each agent with an objective-wise estimate of its own contribution, which can subsequently be combined with the individual preference to guide its policy update.
This provides each agent with an objective-wise estimate of its own contribution, which is subsequently combined with the individual preference to guide its policy update, as illustrated in Figure~\ref{fig:system_overview}.

\subsection{Static Meta-Gradient Informed Preference Adaptation}\label{subsec:static_meta}

The objective-wise counterfactual credit developed above provides each agent with separate estimates of its contribution across the team objectives. The next question is how these objective-wise contributions should be combined when updating the individual policy. 
Rather than fixing this combination manually, we aim to let each agent learn how its individual preference should guide its learning so that the resulting behaviour better contributes to the overall team performance.

This naturally leads to a learning-to-learn problem. In addition to learning the policy itself, the agent must learn how its learning signal should be constructed from the objective-wise credit available to it.
Inspired by meta-gradient reinforcement learning~\citep{xu2018meta-rl}, we develop a meta-gradient preference adaptation mechanism that allows each agent to adapt its individual preference according to how that preference affects subsequent team performance. Specifically, we treat the individual preference as a differentiable meta-variable that controls how the agent's objective-wise credit is combined during its policy update.

For each agent $i$, we introduce an unconstrained meta-parameter $\boldsymbol{\eta}_i\in\mathbb{R}^{d}$ and define its individual preference as
\[
\boldsymbol{\omega}_i
=
\operatorname{softmax}(\boldsymbol{\eta}_i).
\]
This unit-temperature softmax mapping maintains $\boldsymbol{\omega}_i$ on the preference simplex while preserving differentiability with respect to~$\boldsymbol{\eta}_i$.

The centralised vector critic $\mathbf{Q}_{\boldsymbol{\psi}}$ is trained independently using objective-wise TD($\lambda$) targets, without scalarising the objectives using $\boldsymbol{\omega}_i$. Within the computational graph used for the meta-gradient update, the critic update is therefore treated as independent of the preference meta-parameter $\boldsymbol{\eta}_i$, i.e., $\frac{\partial \boldsymbol{\psi}'}{\partial \boldsymbol{\eta}_i}=0$.
Accordingly, for agent $i$ at timestep $t$, the objective-wise counterfactual advantage $\mathbf{A}_{i,t}^{\mathrm{cf}}\in\mathbb{R}^{d}$ computed from the critic is detached and treated as a fixed training signal in the following meta-gradient calculation.

\begin{figure}[t]
    \centering
    \includegraphics[width=0.87\linewidth]{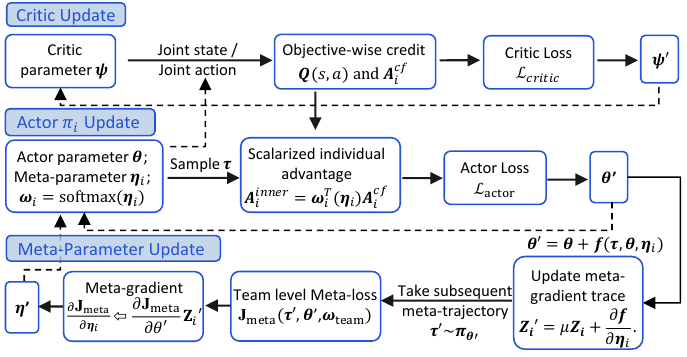}
    \caption{Illustration of training update in~\algoname{}-S.}
    \label{fig:static_meta_gradient_update}
    \Description{This figure shows an illustration of training update in~\algoname{}-S which is also described in Sect.~\ref{subsec:static_meta}}
\end{figure}

\subsubsection{Preference-Guided Actor Update\label{Preference-GuidedAU}}

As illustrated in Figure~\ref{fig:static_meta_gradient_update}, during each policy update, a trajectory
\(\tau=
\left\{\mathbf{o}_t,\mathbf{a}_t,\mathbf{r}_t\right\}_{t=0}^{T-1}
\)
is collected from the current joint policy. 
The objective-wise counterfactual advantages $\mathbf{A}^{\mathrm{cf}}_{i,t}$ used below are then computed from this trajectory using the centralised critic.

For each agent, its current individual preference determines how its objective-wise counterfactual credit is combined into the scalar learning signal used by the actor. Specifically,
\[
A^{\mathrm{inner}}_{i,t}
=
\boldsymbol{\omega}_i^\top
\mathbf{A}^{\mathrm{cf}}_{i,t}
=\operatorname{softmax}(\boldsymbol{\eta}_i)^\top\mathbf{A}^{\mathrm{cf}}_{i,t}.
\]
For agent $i$, the corresponding one-step actor loss is
\[
\ell_{i,t}(\boldsymbol{\theta},\boldsymbol{\eta}_i)=-\log \pi_{\boldsymbol{\theta}}(a_{i,t}\mid h_{i,t},e_i)A^{\mathrm{inner}}_{i,t}.
\]
Over a trajectory of length $T$, the individual actor loss is
\[
J_i(\tau,\boldsymbol{\theta},\boldsymbol{\eta}_i)=\frac{1}{T}\sum_{t=0}^{T-1}\ell_{i,t}(\boldsymbol{\theta},\boldsymbol{\eta}_i),
\]
which gives
\[
J_i(\tau,\boldsymbol{\theta},\boldsymbol{\eta}_i)=
-\frac{1}{T}\sum_{t=0}^{T-1}\log \pi_{\boldsymbol{\theta}}(a_{i,t}\mid h_{i,t},e_i)
\left[
\boldsymbol{\omega}_i^\top\mathbf{A}^{\mathrm{cf}}_{i,t}
\right].
\]

During centralised training, the learning signals from all agents are used jointly to update the shared actor parameters $\boldsymbol{\theta}$. The individual actor losses are therefore averaged to form the inner actor objective,
\[
\mathcal{L}_{\mathrm{actor}}=
J_{\mathrm{inner}}(\tau,\boldsymbol{\theta},\boldsymbol{\eta})
=
\frac{1}{N}\sum_{i=1}^{N}J_i(\tau,\boldsymbol{\theta},\boldsymbol{\eta}_i),
\]
where
\(
\boldsymbol{\eta}
=
\left\{
\boldsymbol{\eta}_1,\ldots,\boldsymbol{\eta}_N
\right\}.
\)
Equivalently,
\begin{equation}
\label{eq:J_inner}
J_{\mathrm{inner}}(\tau,\boldsymbol{\theta},\boldsymbol{\eta})
=
-\frac{1}{NT}\sum_{i=1}^{N}\sum_{t=0}^{T-1} \log \pi_{\boldsymbol{\theta}}(a_{i,t}\mid h_{i,t},e_i)
\left[
\boldsymbol{\omega}_i^\top
\mathbf{A}^{\mathrm{cf}}_{i,t}
\right]
\end{equation}
where $\boldsymbol{\omega}_i=\operatorname{softmax}(\boldsymbol{\eta}_i)$.
The factor $1/N$ simply averages the gradient contributions from the agents and does not require the individual preferences to average to
$\boldsymbol{\omega}_{\mathrm{team}}$.

As shown in Figure~\ref{fig:static_meta_gradient_update}, the actor parameters are updated from this preference-guided inner objective as
\begin{equation}
\label{eq:actor_update}
f_{\boldsymbol{\theta}}(\tau,\boldsymbol{\theta},\boldsymbol{\eta})
=
-\alpha_{\theta}\frac{\partial J_{\mathrm{inner}}(\tau,\boldsymbol{\theta},\boldsymbol{\eta})}
{\partial\boldsymbol{\theta}}
\end{equation}
the updated actor parameters are
\begin{equation}
\label{eq:theta_update}
\boldsymbol{\theta}'
=
\boldsymbol{\theta}
+
f_{\boldsymbol{\theta}}
(\tau,\boldsymbol{\theta},\boldsymbol{\eta}),
\end{equation}
or equivalently,
\[
\boldsymbol{\theta}'
=
\boldsymbol{\theta}-\alpha_{\theta}
\frac{\partial
J_{\mathrm{inner}}
(\tau,\boldsymbol{\theta},\boldsymbol{\eta})}
{\partial\boldsymbol{\theta}}.
\]
Thus, each agent's individual preference determines how its own objective-wise counterfactual credit contributes to the update of the actor.

\subsubsection{Meta-Gradient Preference Update}

The preference-guided actor update above determines how the current individual preferences influence the policy learning.
The remaining question is how these preferences should themselves be adapted. Our goal is to adjust each meta-parameter $\boldsymbol{\eta}_i$ according to whether the actor update induced by its corresponding individual preference leads to improved subsequent team performance.

Following the principle of online cross-validation~\citep{sutton1992adapting}, we evaluate the effect of the actor update on a subsequent sample of experience. After updating the shared actor from $\boldsymbol{\theta}$ to $\boldsymbol{\theta}'$ using the trajectory $\tau$, we collect a subsequent trajectory 
$\tau'=\left\{\mathbf{o}'_t,\mathbf{a}'_t,\mathbf{r}'_t\right\}_{t=0}^{T'-1},  
\tau'\sim\pi_{\boldsymbol{\theta}'}.$
The subsequent trajectory is used to construct a meta-objective that evaluates the updated policy according to the fixed team preference $\boldsymbol{\omega}_{\mathrm{team}}$.

For the subsequent trajectory, we again compute the objective-wise counterfactual advantage for each agent using the centralised vector critic, denoted by $\mathbf{A}_{i,t}^{\mathrm{cf}\prime}\in\mathbb{R}^{d}$.
Unlike the inner actor update, however, these objective-wise contributions are now evaluated according to the fixed team preference rather than the individual preference. We therefore define the team-evaluated meta-advantage for agent $i$ as
\[
A_{i,t}^{\mathrm{meta}}=\boldsymbol{\omega}_{\mathrm{team}}^\top\mathbf{A}_{i,t}^{\mathrm{cf}\prime}.
\]
This provides a scalar measure of whether agent $i$'s action under the updated policy contributes positively to the team objective specified by $\boldsymbol{\omega}_{\mathrm{team}}$.

Using this team-evaluated meta-advantage, we define the meta-objective for the updated actor using the same policy-gradient structure as the actor objective:
\[
J_{\mathrm{meta}}(\tau',\boldsymbol{\theta}',\boldsymbol{\omega}_{\mathrm{team}})
=
-\frac{1}{NT'}\sum_{i=1}^{N}\sum_{t=0}^{T'-1}
\log\pi_{\boldsymbol{\theta}'}(a_{i,t}'\mid h_{i,t}',e_i)A_{i,t}^{\mathrm{meta}}.
\]
The two objectives therefore play different roles in the learning process. $J_{\mathrm{inner}}$ is the primary reinforcement-learning objective used to update the actor parameters $\boldsymbol{\theta}$ under the current individual preferences, whereas $J_{\mathrm{meta}}$ is the meta-learning objective used to adapt the preference meta-parameters $\boldsymbol{\eta}_i$ according to the subsequent team performance under $\boldsymbol{\omega}_{\mathrm{team}}$.

The purpose of the meta-gradient $\frac{\partial J_{\mathrm{meta}}}{\partial\boldsymbol{\eta}_i}$ is therefore to determine how changing agent $i$'s preference meta-parameter $\boldsymbol{\eta}_i$ would affect this subsequent team-level meta-objective $J_{\mathrm{meta}}$. By the chain rule,
\[
\frac{\partial J_{\mathrm{meta}}}{\partial\boldsymbol{\eta}_i}
=
\frac{\partial J_{\mathrm{meta}}}{\partial\boldsymbol{\theta}'}
\frac{d\boldsymbol{\theta}'}{d\boldsymbol{\eta}_i}.
\]

The remaining term $\frac{d\boldsymbol{\theta}'}{d\boldsymbol{\eta}_i}$ describes the sensitivity of the updated actor parameters to agent $i$'s preference meta-parameter. Recall that the actor update is defined in Equation~\ref{eq:theta_update}.
% \[
% \boldsymbol{\theta}'=\boldsymbol{\theta}+f_{\boldsymbol{\theta}}(\tau,\boldsymbol{\theta},\boldsymbol{\eta}).
% \]
Taking the total derivative with respect to $\boldsymbol{\eta}_i$ gives
\[
\frac{d\boldsymbol{\theta}'}{d\boldsymbol{\eta}_i}=
\left(
\mathbf{I}+\frac{\partial f_{\boldsymbol{\theta}}}{\partial\boldsymbol{\theta}}
\right)
\frac{d\boldsymbol{\theta}}{d\boldsymbol{\eta}_i}
+
\frac{\partial f_{\boldsymbol{\theta}}}{\partial\boldsymbol{\eta}_i}.
\]

The term $\frac{\partial f_{\boldsymbol{\theta}}}{\partial\boldsymbol{\theta}}$ is a high-dimensional Jacobian of the actor update and is expensive to compute and propagate explicitly through successive learning updates. 
Following meta-gradient reinforcement learning~\citep{xu2018meta-rl}, we therefore approximate the accumulated sensitivity using a meta-gradient trace
$Z_i \approx \frac{d\boldsymbol{\theta}}{d\boldsymbol{\eta}_i}$,
which is updated as
\begin{equation}
Z_i'=\mu Z_i+\frac{\partial f_{\boldsymbol{\theta}}}{\partial\boldsymbol{\eta}_i}.
\label{mu_Z}
\end{equation}
Here, $\mu\in[0,1]$ controls how much sensitivity from previous actor updates is retained in the trace.
In our experiments, we use $\mu=0$, corresponding to the greedy meta-gradient setting that considers only the effect of the current meta-parameter on the immediately resulting actor update. 

The remaining derivative $\frac{\partial f_{\boldsymbol{\theta}}}{\partial\boldsymbol{\eta}_i}$ can be computed directly from the differentiable dependence of the actor update on the individual preference.
Since the objective-wise counterfactual advantage $\mathbf{A}^{\mathrm{cf}}_{i,t}$ is detached and treated as a fixed training signal, according to actor update Equation~\ref{eq:actor_update}, the sensitivity of the actor update to agent $i$'s preference meta-parameter is
\begin{equation}
\label{eq:f_theta_to_eta_i}
%\begin{aligned}
\frac{\partial f_{\boldsymbol{\theta}}}{\partial\boldsymbol{\eta}_i}
\!=\!
-\alpha_{\theta}
\frac{\partial}{\partial\boldsymbol{\eta}_i}
%\left(
\frac{\partial J_{\mathrm{inner}}}{\partial\boldsymbol{\theta}}%\right)
\!=\!
\frac{\alpha_{\theta}}{NT}\!
\sum_{t=0}^{T-1}\!\nabla_{\boldsymbol{\theta}}\log\pi_{\boldsymbol{\theta}}(a_{i,t}\!\mid\! h_{i,t},e_i)\!
\left[
\left(
\frac{\partial\boldsymbol{\omega}_i}{\partial\boldsymbol{\eta}_i}
\right)^\top\!\!\!
\mathbf{A}^{\mathrm{cf}}_{i,t}\!
\right]^\top\!\!\!.
%\end{aligned}
\end{equation}

The preference Jacobian $\frac{\partial\boldsymbol{\omega}_i}{\partial\boldsymbol{\eta}_i}$ appearing in this expression can be computed analytically, since the 
individual preference depends on the meta-parameter through $\boldsymbol{\omega}_i(\boldsymbol{\eta}_i)=\operatorname{softmax}(\boldsymbol{\eta}_i)$.
The corresponding softmax Jacobian is
\[
\frac{\partial\boldsymbol{\omega}_i}{\partial\boldsymbol{\eta}_i}
=\operatorname{diag}(\boldsymbol{\omega}_i)-\boldsymbol{\omega}_i\boldsymbol{\omega}_i^\top.
\]

With the greedy meta-gradient setting $\mu=0$, the trace (\ref{mu_Z}) becomes $Z_i'=\frac{\partial f_{\boldsymbol{\theta}}}{\partial\boldsymbol{\eta}_i}$.

The meta-gradient for agent $i$ can therefore be approximated as
\begin{equation}
\label{eq:meta_gradient}
\frac{\partial J_{\mathrm{meta}}}{\partial\boldsymbol{\eta}_i}
\approx
\frac{\partial J_{\mathrm{meta}}}{\partial\boldsymbol{\theta}'}Z_i'
=
\frac{\partial J_{\mathrm{meta}}}{\partial\boldsymbol{\theta}'}\frac{\partial f_{\boldsymbol{\theta}}}{\partial\boldsymbol{\eta}_i}.
\end{equation}
This gradient captures how changing agent $i$'s individual preference affects the subsequent team-level meta-objective through its influence on the actor update. 

The resulting meta-gradients are used to update the preference meta-parameters. In our implementation, the meta-parameters $\boldsymbol{\eta}=\{\boldsymbol{\eta}_1,\ldots,\boldsymbol{\eta}_N\}$ are optimised using Adam with meta learning rate $\beta_{\omega}$.
After the meta-parameter update, the corresponding individual preferences are obtained through the softmax mapping as
\(
\boldsymbol{\omega}_i'=\operatorname{softmax}(\boldsymbol{\eta}_i').
\)

Although the individual preferences can be adapted through the meta-gradient learning process, each agent is still represented by a single meta-parameter vector $\boldsymbol{\eta}_i$ throughout the interaction. 
Consequently, the resulting preference $\boldsymbol{\omega}_i$ is state-independent: it can capture a different overall objective emphasis for each agent, but cannot change in response to the agent's current observation, behaviour, or cooperative situation.
% We refer to this formulation as the static variant of our method, named \algoname{}-S. Details of \algoname{}-S is provided in Algorithm~\ref{app:algorithm_s} in Appendix~\ref{app:algo}.
We refer to this static formulation as \algoname{}-S, with details provided in Algorithm~\ref{app:algorithm_s} in Appendix~\ref{app:algo}.

This static formulation therefore provides agent-specific preference adaptation at the training level, 
% but does not yet provide the situational adaptability motivated in Section~\ref{subsec:MOMARL_modelling}. 
but does not yet allow individual preferences to adapt to the current cooperative situation.
In cooperative tasks, however, the role or contribution required from an agent may change as the interaction evolves. 
This motivates extending the preference adaptation mechanism from a single learned preference for each agent to preferences that can dynamically vary according to the current cooperative context.

\subsection{Dynamic Meta-Gradient Informed Preference Adaptation}\label{subsec:dynamic_meta}

To enable individual preferences to vary with the current cooperative situation, we replace the single state-independent meta-parameter $\boldsymbol{\eta}_i$ used for each agent in the static formulation with a differentiable preference controller $g_{\boldsymbol{\phi}}$. Rather than learning one fixed preference representation for agent $i$, the controller generates a preference meta-variable from the agent's current local representation at each timestep. This dynamic preference pathway is illustrated in Figure~\ref{fig:system_overview}.
Specifically,
\[
\boldsymbol{\eta}_{i,t}=g_{\boldsymbol{\phi}}(\mathbf{h}_{i,t}),
\]
and the corresponding dynamic individual preference is
\[
\boldsymbol{\omega}_{i,t}=\operatorname{softmax}(\boldsymbol{\eta}_{i,t})
=
\operatorname{softmax}\left(g_{\boldsymbol{\phi}}(\mathbf{h}_{i,t})\right).
\]
Here, $\boldsymbol{\phi}$ denotes the learnable parameters of the preference controller, and $\mathbf{h}_{i,t}$ denotes the local recurrent representation available to agent $i$ at timestep $t$. Conditioning the controller on $\mathbf{h}_{i,t}$ allows the resulting preference to vary across agents and timesteps according to the current cooperative situation.

The dynamically generated preference is also explicitly provided to the actor. Specifically, the policy is conditioned jointly on the local recurrent representation and the current individual preference,
$a_{i,t}\sim
\pi_{\boldsymbol{\theta}}\left(\cdot\mid\mathbf{h}_{i,t},\boldsymbol{\omega}_{i,t}
\right)$.
This allows changes in the generated preference to directly influence the agent's behaviour at each timestep.

The same dynamic preference is used to scalarise the objective-wise counterfactual credit during policy learning:
\[
A_{i,t}^{\mathrm{inner,dyn}}
=
\boldsymbol{\omega}_{i,t}^\top\mathbf{A}_{i,t}^{\mathrm{cf}}.
\]

The subsequent actor update and meta-gradient preference update follow the same formulation as in the static case, with the dynamically generated $\boldsymbol{\omega}_{i,t}$ replacing the static $\boldsymbol{\omega}_i$. The centralised vector critic and objective-wise counterfactual credit assignment remain unchanged.
This dynamic formulation defines the second variant of our method, \algoname{}-D, with details provided in Algorithm~\ref{app:algorithm_d} in Appendix~\ref{app:algo}.

\section{Experiments}
\subsection{Experiment Setup}

\textbf{Benchmark Problems.}
We conduct experiments using the Multi-Agent Particle Environment (MPE)~\citep{lowe2017multi}, which is a widely used RL benchmark. 
We extend the original MPE to a multi-objective setting by retaining the original task reward as the first objective and introducing a second objective that captures the energy consumed during interaction. 
We evaluate the proposed methods and baselines on four cooperative tasks: \textit{Spread}, \textit{Formation}, \textit{Line}, and \textit{Reference}. 
The first three tasks involve three agents with different spatial coordination requirements, while \textit{Reference} involves two agents and additionally requires communication for successful cooperation. 
Full environment definitions are provided in Appendix~\ref{app:benchmarks}.

\textbf{Evaluation Metrics.}
We evaluate the quality of the resulting Pareto fronts using three standard metrics: Hypervolume (HV), Expected Utility (EU), and Sparsity (SP), following the formalism in~\citep{zitzler1999multiobjective, zintgraf2015quality, hayes2022practical}.
HV measures the overall quality and coverage of the Pareto front, EU evaluates performance across different objective preferences, and SP measures the distribution of solutions along the front. 
Higher HV and EU and lower SP indicate better performance. 
Since SP is scale-dependent and less directly related to decision quality, we report it mainly as a complementary metric to assess coverage diversity.
More details can be found in Appendix~\ref{app:eval_metrics}.

\textbf{Baselines.}
As dedicated MOMARL algorithms remain relatively limited, we consider five baselines of increasing complexity: \textit{DTDE-PPO}, \textit{CTDE-PPO-V}, \textit{CTDE-PPO-Q}, \textit{MO-COMA}, and \textit{MO-MIX}.
\textit{DTDE-PPO} represents the fully decentralised setting, where each agent independently learns its policy using PPO~\citep{schulman2017proximal-PPO} network. 
Given the surprisingly strong performance of PPO previously observed in cooperative MARL~\citep{yu2022MAPPO}, we further extend MAPPO to the MOMARL setting under CTDE and consider two variants, \textit{CTDE-PPO-V} and \textit{CTDE-PPO-Q}, using centralised $V$- and $Q$-critics, respectively.
\textit{MO-COMA} extends COMA~\citep{foerster2018counterfactual} to the multi-objective setting and provides counterfactual credit assignment for individual agents across multiple objectives. It also serves as an ablation baseline for evaluating the additional preference-adaptation mechanisms introduced in our methods.
We additionally compare against \textit{MO-MIX}~\citep{hu2023momix}, a dedicated MOMARL algorithm based on multi-objective value decomposition.
All methods except MO-MIX operate in continuous action spaces. Since MO-MIX only supports discrete actions, it is evaluated using a discretised version of the action space.
Further implementation details are provided in Appendix~\ref{app:training_detail}.

\subsection{Results}

We now present the empirical results across four benchmark problems.
All methods are trained under the same total interaction budget of $5\times10^{6}$ joint environment steps per environment.
Full training settings and hyperparameters are provided in Appendix~\ref{app:training_detail}. 

Figure~\ref{fig:pareto_fronts} compares the Pareto fronts obtained by the different methods, while Table~\ref{tab:eval_results} reports the quantitative evaluation results as mean $\pm$ standard deviation.
Each marker corresponds to a non-dominated solution in the estimated Pareto front. The connecting lines are included only as a visual guide between neighbouring solutions.
Across the benchmark problems, the proposed \algoname{} variants consistently achieve strong Pareto-front quality, with \algoname{}-D showing the best overall performance. 
Further analysis of preference adaptation is provided in Appendix~\ref{app:additional_exp}.

\begin{table*}[t]
    \centering
    \caption{Evaluation of the Pareto front quality.}
    \label{tab:eval_results}
      \scalebox{0.88}{
  \begin{tabular}{l| c| ccccc cc}
    \toprule
    \multirow{2}{*}{Environment} 
      & \multirow{2}{*}{Metric} 
      & \multicolumn{7}{c}{Method} \\
    \cmidrule(lr){3-9}
     &  & DTDE-PPO & CTDE-PPO-V & CTDE-PPO-Q & MO-COMA & MO-MIX & \algoname-S & \algoname-D \\
    \midrule
    \multirow{3}{*}{Spread} & HV & 1222.55 $\pm$ 15.43 & 1434.46 $\pm$ 52.34 & 1512.17 $\pm$ 19.41 & 1570.08 $\pm$ 8.61 & 1213.48 $\pm$ 131.15 & 1581.40 $\pm$ 30.05 & \textbf{1594.55 $\pm$ 20.34} \\
     & EU & 36.57 $\pm$ 0.24 & 38.16 $\pm$ 0.54 & 39.32 $\pm$ 0.12 & 39.85 $\pm$ 0.08 & 36.44 $\pm$ 2.26 & 40.02 $\pm$ 0.32 & \textbf{40.14 $\pm$ 0.25} \\
     & SP & 60.58 $\pm$ 76.77 & 14.03 $\pm$ 3.08 & 2.86 $\pm$ 2.32 & 1.20 $\pm$ 0.82 & \textbf{0.07 $\pm$ 0.07} & 1.39 $\pm$ 1.06 & 12.08 $\pm$ 19.77 \\
    \midrule
    \multirow{3}{*}{Formation} & HV & 1359.53 $\pm$ 10.20 & 1815.27 $\pm$ 40.36 & 1975.36 $\pm$ 40.19 & 2178.79 $\pm$ 23.31 & 1525.67 $\pm$ 333.53 & 2187.33 $\pm$ 7.91 & \textbf{2196.13 $\pm$ 15.76} \\
     & EU & 37.79 $\pm$ 0.25 & 42.59 $\pm$ 0.34 & 44.02 $\pm$ 0.31 & 45.76 $\pm$ 0.20 & 39.21 $\pm$ 4.77 & 45.85 $\pm$ 0.19 & \textbf{46.03 $\pm$ 0.25} \\
     & SP & 918.66 $\pm$ 526.72 & 1.71 $\pm$ 0.80 & 3.53 $\pm$ 1.49 & 2.14 $\pm$ 1.37 & \textbf{0.07 $\pm$ 0.07} & 0.77 $\pm$ 0.22 & 1.52 $\pm$ 0.45 \\
    \midrule
    \multirow{3}{*}{Line} & HV & 1245.48 $\pm$ 70.17 & 1697.30 $\pm$ 13.68 & 1696.68 $\pm$ 8.88 & 1958.71 $\pm$ 26.19 & 1239.09 $\pm$ 433.07 & 1957.49 $\pm$ 22.05 & \textbf{1965.89 $\pm$ 24.11} \\
     & EU & 36.54 $\pm$ 0.53 & 41.19 $\pm$ 0.07 & 41.25 $\pm$ 0.28 & 43.64 $\pm$ 0.33 & 35.32 $\pm$ 7.22 & 43.71 $\pm$ 0.26 & \textbf{43.74 $\pm$ 0.26} \\
     & SP & 96.38 $\pm$ 134.45 & 33.32 $\pm$ 27.87 & 14.65 $\pm$ 4.24 & 0.86 $\pm$ 0.57 & \textbf{0.12 $\pm$ 0.09} & 0.81 $\pm$ 0.62 & 0.79 $\pm$ 0.38 \\
    \midrule
    \multirow{3}{*}{Reference} & HV & 1843.75 $\pm$ 24.38 & 2082.20 $\pm$ 28.47 & 2413.35 $\pm$ 36.78 & 2567.07 $\pm$ 21.76 & 2016.33 $\pm$ 69.76 & 2587.41 $\pm$ 28.06 & \textbf{2587.44 $\pm$ 27.99} \\
     & EU & 44.64 $\pm$ 0.17 & 46.93 $\pm$ 0.24 & 48.09 $\pm$ 0.51 & 50.67 $\pm$ 0.18 & 46.02 $\pm$ 0.87 & 50.81 $\pm$ 0.14 & \textbf{50.85 $\pm$ 0.21} \\
     & SP & -- & 1254.31 $\pm$ 12.10 & 107.04 $\pm$ 42.06 & 0.90 $\pm$ 0.53 & 3.08 $\pm$ 3.04 & 0.62$\pm$0.11 & \textbf{0.62 $\pm$ 0.09} \\
    \bottomrule
  \end{tabular}
}
\end{table*}

\begin{figure}[t]
    \centering
    \begin{subfigure}[t]{0.23\textwidth}
        \centering
        \includegraphics[width=\linewidth]{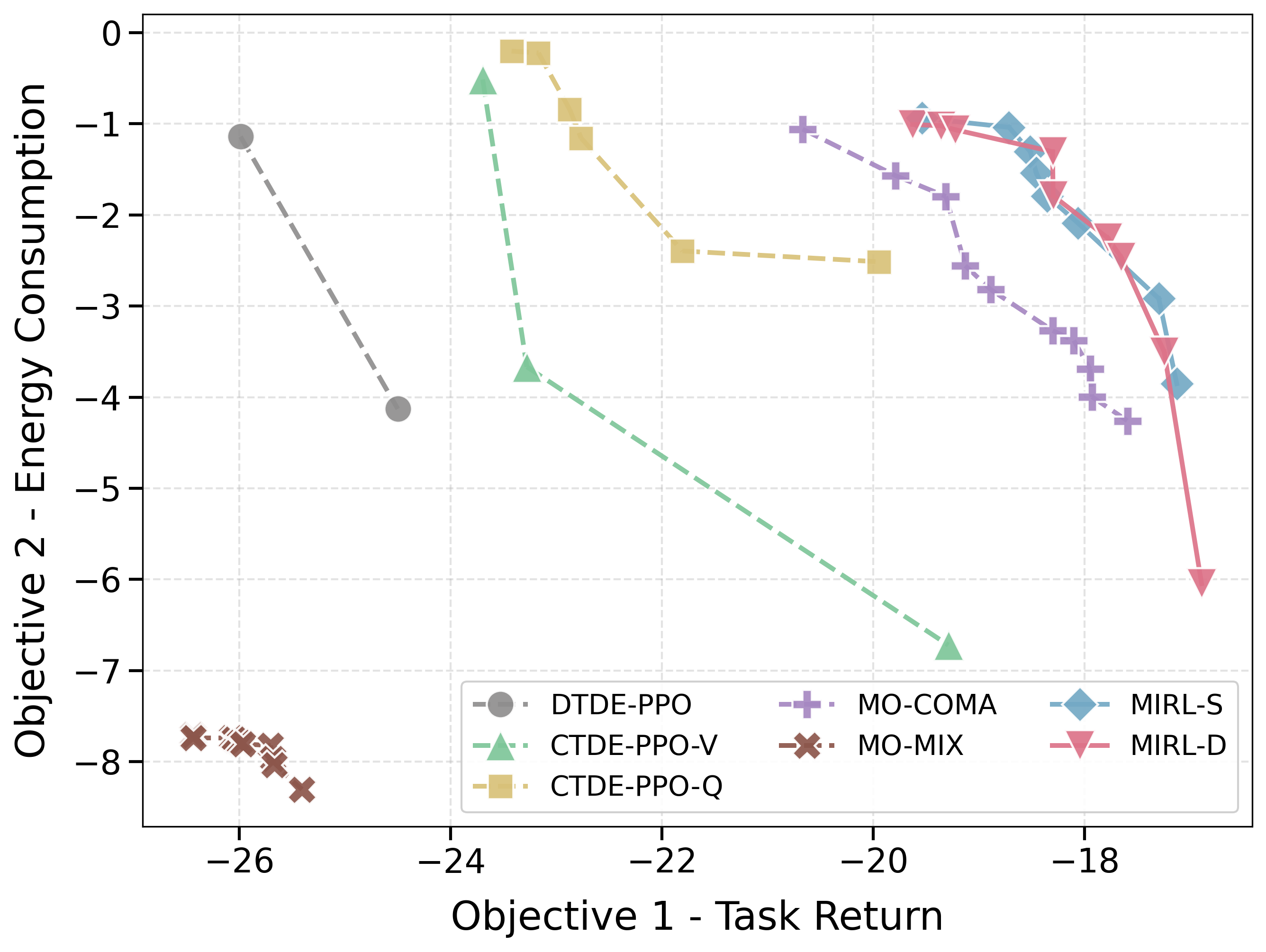}
        \caption{Spread.}
        \label{fig:pareto_spread}
    \end{subfigure}
    \hfill
    \begin{subfigure}[t]{0.23\textwidth}
        \centering
        \includegraphics[width=\linewidth]{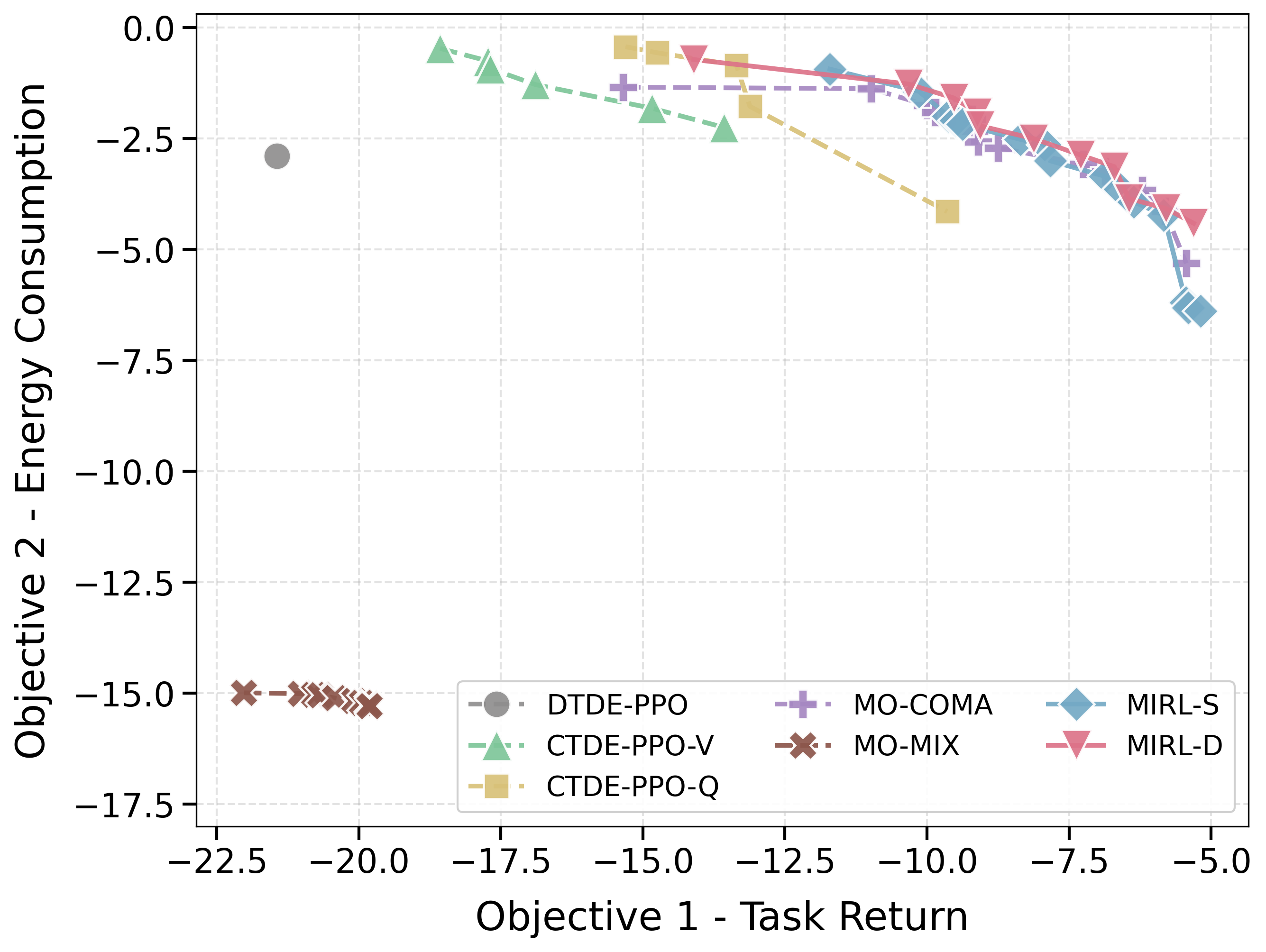}
        \caption{Formation.}
        \label{fig:pareto_formation}
    \end{subfigure}
    \\
    \begin{subfigure}[t]{0.23\textwidth}
        \centering
        \includegraphics[width=\linewidth]{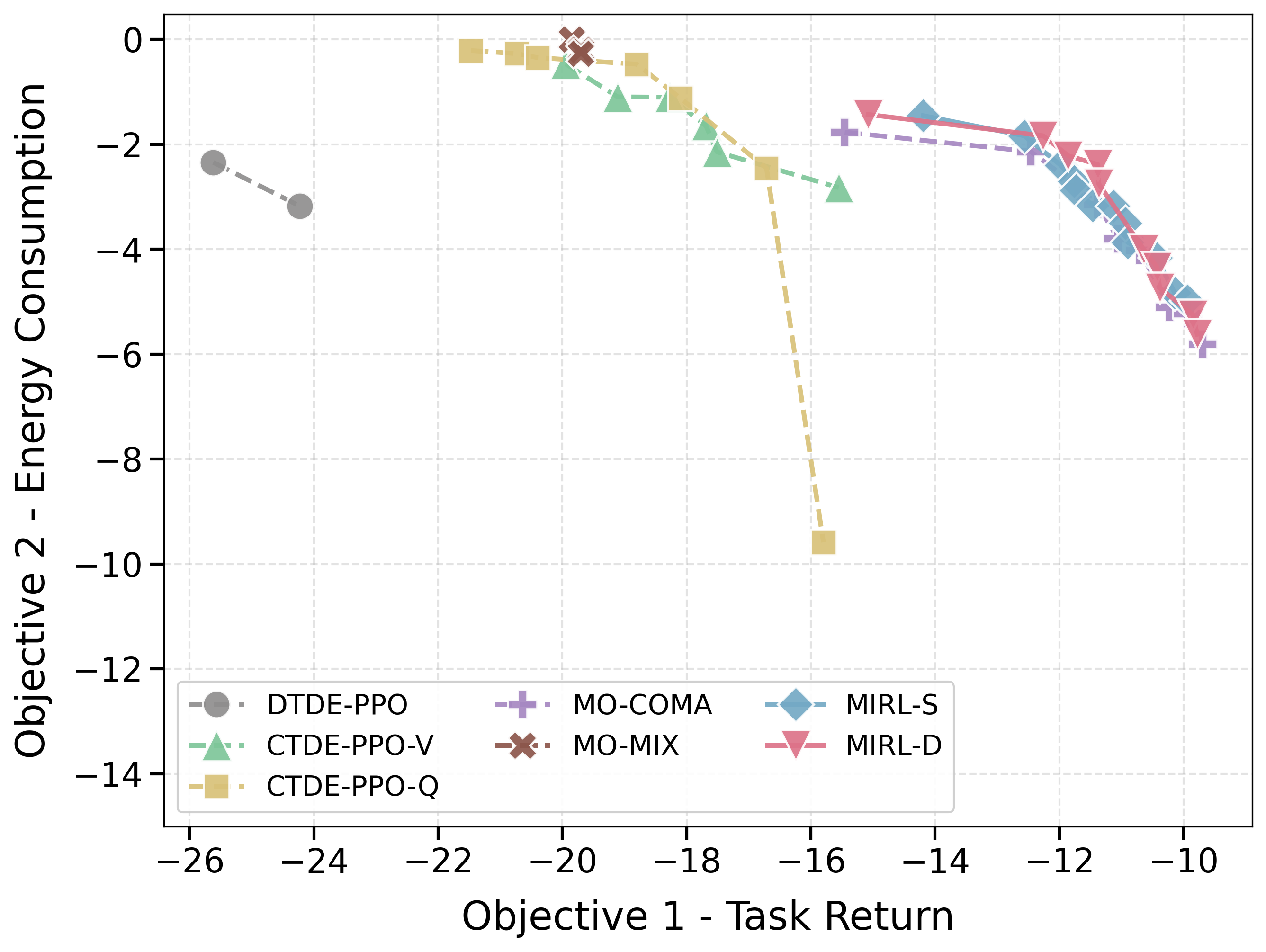}
        \caption{Line.}
        \label{fig:pareto_line}
    \end{subfigure}
    \hfill
    \begin{subfigure}[t]{0.23\textwidth}
        \centering
        \includegraphics[width=\linewidth]{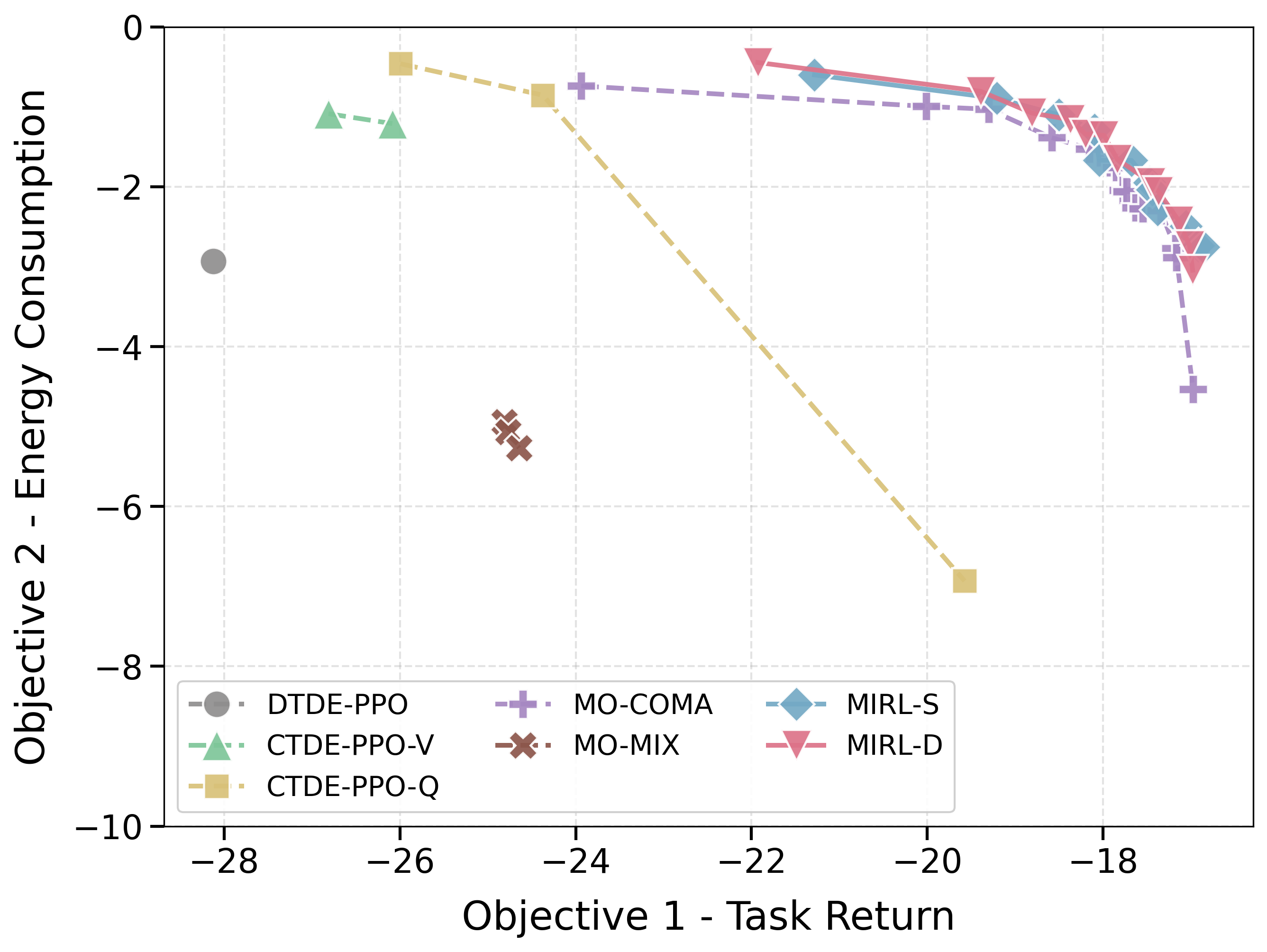}
        \caption{Reference.}
        \label{fig:pareto_reference}
    \end{subfigure}
    \Description{Pareto front comparison shows that the proposed algorithm variant MIRL-S is competitive in comparison with the listed algorithms on four different benchmark problems, while the dynamic variant MIRL-D has a small advantage as also clear from the information in Tabl.~\ref{tab:eval_results}.}
    \caption{Pareto front comparison of the \algoname{} variants with the baselines across four benchmark problems.
    }
    \label{fig:pareto_fronts}
\end{figure}

Among the PPO-based baselines, DTDE-PPO consistently shows the weakest performance, illustrating the difficulty of learning cooperative behaviours through fully decentralised training in these tasks.
Introducing centralised information during training substantially improves both Pareto-front quality and coverage.
In particular, CTDE-PPO-Q generally performs better than CTDE-PPO-V, suggesting that conditioning the critic on the joint action provides a more informative learning signal for coordinating individual actions.
It is worth noting that these PPO-based methods constitute strong rather than trivial baselines. PPO-based methods have previously demonstrated surprisingly strong performance in MARL~\citep{yu2022MAPPO}, and the variants here retain this strong policy-optimisation backbone while extending it to the multi-objective setting.

MO-MIX exhibits a different behaviour from the other methods.
As shown in Figure~\ref{fig:pareto_fronts}, it often produces relatively dense groups of solutions, which is also reflected by its low SP values, but these solutions tend to occupy a more limited region of the objective space and result in substantially lower HV and EU.
One possible reason is that MO-MIX is a single preference-conditioned network, which may be more sensitive to hyperparameter choices and can be more difficult to train.
Under the same overall interaction budget, a single model must represent policies across the full preference space, and this shared representation may, to some extent, compromise performance.
In addition, MO-MIX operates on discrete actions and therefore requires discretisation of the originally continuous MPE action space, which may further limit its achievable performance.

MO-COMA achieves substantially stronger performance than the PPO-based baselines across all tasks, showing that objective-wise counterfactual credit assignment alone already provides an effective learning signal for cooperative MOMARL.
Its performance is particularly strong on \textit{Formation} and \textit{Line}, where it is already close to both \algoname{} variants.
On \textit{Spread} and \textit{Reference}, however, the \algoname~methods provide clearer improvements over MO-COMA. 
We analyse these differences further in the ablation study.

\subsection{Ablation and Further Analysis}
To isolate the contribution of the proposed meta-gradient informed preference adaptation mechanism, we compare the ablation baseline MO-COMA with static \algoname-S and dynamic \algoname-D methods.

The benefit of preference adaptation varies with the coordination requirements of the task.
On \textit{Spread}, \algoname{}-S improves over MO-COMA, and \algoname{}-D provides a further improvement.
This task requires agents to continuously adjust their relative positions and contributions during interaction, making the ability to adapt individual preferences to the current cooperative context particularly useful.
On \textit{Reference}, both \algoname{}-S and \algoname{}-D improve over MO-COMA, while the difference between the two \algoname{} variants is comparatively small. In this task, the cooperative requirement, particularly the communication component, is relatively stable and less dependent on the changing cooperative situation.
The additional flexibility of dynamic preference adaptation therefore provides a smaller advantage.
By contrast, on \textit{Formation} and \textit{Line}, MO-COMA already achieves performance very close to the two \algoname{} variants.
These tasks have relatively simple and stable coordination requirements, and objective-wise counterfactual credit assignment alone appears sufficient to produce strong cooperative behaviour, leaving less room for additional gains from preference adaptation.
% To further examine whether this small gap is related to the limited coordination difficulty of these tasks, we additionally evaluate \textit{Formation} and \textit{Line} with five agents.
% Increasing the number of agents makes the coordination problem more challenging while preserving the underlying task structure. The corresponding results are reported in Appendix~\ref{app:additional_exp}.

\section{Conclusion}
In this work, we show that the multi-objective structure can be exploited to facilitate multi-agent cooperation, rather than being treated solely as an additional optimisation challenge in MOMARL.
Building on this novel perspective, we formulate team-level and individual-level preferences to introduce adaptive agent-level guidance while preserving the system-level objective trade-off.
To enable preference-guided cooperation across agents, we derive objective-wise counterfactual credit assignment together with a meta-gradient informed preference adaptation mechanism, providing a principled way to learn how individual preferences should shape policy updates.
These developments lead to \algoname{}, with static (\algoname{}-S) and dynamic (\algoname{}-D) variants, which achieve strong Pareto-front performance across the benchmark tasks, with \algoname{}-D showing the best overall results.
Our ablation results further demonstrate that meta-gradient informed preference adaptation provides additional benefits beyond counterfactual credit assignment, particularly when cooperation requires greater adaptability.

%%%%%%%%%%%%%%%%%%%%%%%%%%%%%%%%%%%%%%%%%%%%%%%%%%%%%%%%%%%%%%%%%%%%%%%%

%%% The acknowledgments section is defined using the "acks" environment
%%% (rather than an unnumbered section). The use of this environment 
%%% ensures the proper identification of the section in the article 
%%% metadata as well as the consistent spelling of the heading.

\begin{acks}
This work was supported by the United Kingdom Research and Innovation (grant EP/S023208/1), UKRI Centre for Doctoral Training in Robotics and Autonomous Systems at the University of Edinburgh, School of Informatics.
\end{acks}

%%%%%%%%%%%%%%%%%%%%%%%%%%%%%%%%%%%%%%%%%%%%%%%%%%%%%%%%%%%%%%%%%%%%%%%%

%%% The next two lines define, first, the bibliography style to be 
%%% applied, and, second, the bibliography file to be used.

%\newpage

\bibliographystyle{ACM-Reference-Format} 
\bibliography{refs}

% \clearpage
\appendix
\section{Algorithm}\label{app:algo}

Algorithms~\ref{app:algorithm_s} and~\ref{app:algorithm_d} describe MIRL-S and MIRL-D, respectively. 
The total training budget is divided equally among $M$ fixed team preferences, and one policy is trained independently for each preference. MIRL-S learns static agent preferences, whereas MIRL-D generates dynamic preferences using a shared controller. After training, only the non-dominated policies are retained in $\Pi$.

\begin{algorithm}
    \caption{\algoname{}-S}
    \label{app:algorithm_s}
    \begin{algorithmic}[1]
        \Require 
        \\Fixed user-specified team preferences
        $\{\boldsymbol{\omega}_{\mathrm{team},m}\}_{m=1}^{M}$
        evenly spanning the preference space;\\
        actor $\pi_{\boldsymbol{\theta}}$ and centralised vector
        critic $Q_{\boldsymbol{\psi}}$;\\
        actor, critic, and meta learning rates
        $\alpha_{\theta}$, $\alpha_{\psi}$, and $\beta_{\omega}$;\\
        discount factor $\gamma$ and TD trace parameter $\lambda$;\\
        counterfactual sample count $K$ and meta-trace decay $\mu$;\\
        total training budget $B_{\mathrm{total}}$
        
        \State \textbf{Initialisation:}
        \State Divide the total budget evenly:
        $B_m\gets B_{\mathrm{total}}/M$ for all $m$
        \State Initialise the output policy set
        $\Pi\gets\varnothing$

        \For{$m=1,\ldots,M$}
            \State Set
            $\boldsymbol{\omega}_{\mathrm{team}}
            \gets\boldsymbol{\omega}_{\mathrm{team},m}$
            \State Initialise $\boldsymbol{\theta}$ and
            $\boldsymbol{\psi}$
            \State Initialise $\boldsymbol{\eta}_i$ that
            $\boldsymbol{\omega}_i
            =\operatorname{softmax}(\boldsymbol{\eta}_i)
            \approx\boldsymbol{\omega}_{\mathrm{team}}$
            for all agents $i$
            \State Initialise $Z_i\gets0$ for all $i$

            \While{the training steps are fewer than $B_m$}
                \State Collect an inner trajectory
                $\tau\sim\pi_{\boldsymbol{\theta}}$

                \Statex
                \hspace{\dimexpr\algorithmicindent\relax}
                \textbf{Critic update}
                \State Update $\boldsymbol{\psi}$ using
                objective-wise TD($\lambda$) targets from $\tau$
                \State Compute
                $\{\mathbf A^{\mathrm{cf}}_{i,t}\}$
                using $K$ counterfactual samples

                \Statex
                \hspace{\dimexpr\algorithmicindent\relax}
                \textbf{Actor update}
                \State Set
                $\boldsymbol{\omega}_i
                \gets\operatorname{softmax}(\boldsymbol{\eta}_i)$
                for all $i$
                \State Form
                $J_{\mathrm{inner}}
                (\tau,\boldsymbol{\theta},\boldsymbol{\eta})$
                \State 
                $\boldsymbol{\theta}'
                \gets\boldsymbol{\theta}
                +f_{\boldsymbol{\theta}}
                (\tau,\boldsymbol{\theta},\boldsymbol{\eta})$

                \Statex
                \hspace{\dimexpr\algorithmicindent\relax}
                \textbf{Meta-parameter update}
                \State Collect a subsequent trajectory
                $\tau'\sim\pi_{\boldsymbol{\theta}'}$
                \State Compute 
                $\{\mathbf A^{\mathrm{cf}\prime}_{i,t}\}$
                from $\tau'$
                \State Form
                $J_{\mathrm{meta}}
                (\tau',\boldsymbol{\theta}',
                \boldsymbol{\omega}_{\mathrm{team}})$
                \State Update $\{Z_i\}_{i=1}^{N}$ and
                $\boldsymbol{\eta}$ using the meta-gradient
                \State Set
                $\boldsymbol{\theta}\gets\boldsymbol{\theta}'$
            \EndWhile

            \State Store             $\pi_m\gets\pi_{\boldsymbol{\theta}}$
            and set $\Pi\gets\Pi\cup\{\pi_m\}$
        \EndFor

        \State 
        $\Pi\leftarrow$ non‐dominated subset of $\Pi$
        \State \Return $\Pi$
    \end{algorithmic}
\end{algorithm}

\begin{algorithm}
    \caption{\algoname{}-D}
    \label{app:algorithm_d}
    \begin{algorithmic}[1]
        \Require
        \\Fixed user-specified team preferences
        $\{\boldsymbol{\omega}_{\mathrm{team},m}\}_{m=1}^{M}$
        evenly spanning the preference space;\\
        actor $\pi_{\boldsymbol{\theta}}$, dynamic preference
        controller $g_{\boldsymbol{\phi}}$, and centralised vector
        critic $Q_{\boldsymbol{\psi}}$;\\
        actor, critic, and meta learning rates
        $\alpha_{\theta}$, $\alpha_{\psi}$, and $\beta_{\omega}$;\\
        discount factor $\gamma$ and TD trace parameter $\lambda$;\\
        counterfactual sample count $K$ and meta-trace decay $\mu$;\\
        total training budget $B_{\mathrm{total}}$

        \State \textbf{Initialisation:}
        \State Divide the total budget evenly:
        $B_m\gets B_{\mathrm{total}}/M$ for all $m$
        \State Initialise the output policy set
        $\Pi\gets\varnothing$

        \For{$m=1,\ldots,M$}
            \State Set
            $\boldsymbol{\omega}_{\mathrm{team}}
            \gets\boldsymbol{\omega}_{\mathrm{team},m}$
            \State Initialise $\boldsymbol{\theta}$,            $\boldsymbol{\psi}$ and $\boldsymbol{\phi}$
            \State Initialise $Z_{\phi}\gets0$

            \While{the training steps are fewer than $B_m$}
                \State Collect an inner trajectory $\tau$ using
                $\pi_{\boldsymbol{\theta}}$ and
                $g_{\boldsymbol{\phi}}$

                \Statex
                \hspace{\dimexpr\algorithmicindent\relax}
                \textbf{Critic update}
                \State Update $\boldsymbol{\psi}$ using
                objective-wise TD($\lambda$) targets from $\tau$
                \State Compute
                $\{\mathbf A^{\mathrm{cf}}_{i,t}\}$
                using $K$ counterfactual samples

                \Statex
                \hspace{\dimexpr\algorithmicindent\relax}
                \textbf{Actor update}
                \State 
                $\boldsymbol{\eta}_{i,t}
                \gets g_{\boldsymbol{\phi}}(\mathbf h_{i,t})$
                and
                $\boldsymbol{\omega}_{i,t}
                \gets\operatorname{softmax}
                (\boldsymbol{\eta}_{i,t})$
                \State Form
                $J_{\mathrm{inner}}
                (\tau,\boldsymbol{\theta},\boldsymbol{\phi})$
                \State
                $\boldsymbol{\theta}'
                \gets\boldsymbol{\theta}
                +f_{\boldsymbol{\theta}}
                (\tau,\boldsymbol{\theta},\boldsymbol{\phi})$

                \Statex
                \hspace{\dimexpr\algorithmicindent\relax}
                \textbf{Meta-parameter update}
                \State Collect a subsequent trajectory $\tau'$ using
                $\pi_{\boldsymbol{\theta}'}$ and
                $g_{\boldsymbol{\phi}}$
                \State Compute
                $\{\mathbf A^{\mathrm{cf}\prime}_{i,t}\}$
                from $\tau'$
                \State Form
                $J_{\mathrm{meta}}
                (\tau',\boldsymbol{\theta}',
                \boldsymbol{\omega}_{\mathrm{team}})$
                \State Update $Z_{\phi}$ and
                $\boldsymbol{\phi}$ using the meta-gradient
                \State Set
                $\boldsymbol{\theta}\gets\boldsymbol{\theta}'$
            \EndWhile

            \State 
            $\pi_m\gets
            \pi_{\boldsymbol{\theta}}$
            and set $\Pi\gets\Pi\cup\{\pi_m\}$
        \EndFor

        \State
        $\Pi\leftarrow$ non-dominated subset of $\Pi$
        \State \Return $\Pi$
    \end{algorithmic}
\end{algorithm}

\section{Experiment Setup Details}\label{app:exp_details}

\subsection{Benchmarks}\label{app:benchmarks}

We evaluate all methods in cooperative tasks derived from the Multi Particle Environments (MPE)~\citep{lowe2017multi,mordatch2017emergence}, using the maintained MPE2/PettingZoo implementation~\citep{terry2021pettingzoo}. MPE models agents and landmarks as particles moving in a two-dimensional continuous space. We consider four tasks: \textsc{Spread}, \textsc{Formation}, \textsc{Line}, and \textsc{Reference}. 
The first three tasks contain three agents, while \textsc{Reference} contains two agents and an explicit communication channel.
Each episode lasts for \(H=25\) environment steps.

We extend each task into a two-objective multi-agent environment. The reward vector of agent \(i\) at timestep \(t\) is
\[
    \mathbf{r}_{i,t} = 
        \left[ r^{\mathrm{task}}_{i,t}, r^{\mathrm{energy}}_{i,t} \right]
    = \left[r^{\mathrm{task}}_{i,t} , -E_{i,t} \right],
\]
where \(r^{\mathrm{task}}_{i,t}\) is the task reward provided by the underlying MPE scenario and \(E_{i,t}\) is the energy consumed by the agent. Both objectives are therefore maximised: the first encourages task completion, whereas the second encourages energy-efficient behaviour.

% For evaluation, we report the undiscounted team return
% \[
%     \mathbf{R}    =
%     \sum_{t=0}^{H-1} \frac{1}{N} \sum_{i=1}^{N} \mathbf{r}_{i,t},
% \]
% where \(N\) is the number of agents. 

At each timestep, we define the team reward vector as the average of the per-agent reward vectors,
\(
\overline{\mathbf{r}}_{t}=\frac{1}{N}\sum_{i=1}^{N}\mathbf{r}_{i,t},
\)
where \(N\) is the number of agents. 
Averaging prevents the shared MPE task reward from being counted repeatedly and provides a consistent reward scale across environments containing different numbers of agents.
The discounted team return over an episode is
\[
\mathbf{G}=\sum_{t=0}^{H-1}\gamma^t\overline{\mathbf{r}}_{t},
\]
where \(\gamma=0.99\) across all methods. 
The expected discounted team returns are used to construct the empirical Pareto fronts and calculate the evaluation metrics.

For all continuous-control methods, each agent selects a two-dimensional movement command
\[
    \mathbf{f}_{i,t} = \left[
        f^x_{i,t} , f^y_{i,t}
    \right]^{\top}
    \in[-1,1]^2.
\]
This command is mapped to the native MPE control interface according to
\[
    \mathbf{u}_{i,t} = \left[
        0,\,
        [-f^x_{i,t}]_+,\,
        [f^x_{i,t}]_+,\,
        [-f^y_{i,t}]_+,\,
        [f^y_{i,t}]_+
    \right]^{\top},
\]
where \([x]_+=\max(x,0)\). 
The learned movement action is thus two-dimensional; the five-dimensional vector above is only the internal representation expected by MPE.

We define movement energy as the squared control effort
\[
    E^{\mathrm{move}}_{i,t}  =
    \alpha \left\lVert \mathbf{f}_{i,t}\right\rVert_2^2,
\]
where \(\alpha=1\) in all experiments.

\paragraph{\textbf{\textsc{Spread}.}}
The \textsc{Spread} task contains \(N=3\) agents and an equal number of landmarks. The agents must collectively cover all landmarks while avoiding collisions. Let \(\mathbf{p}_{i,t}\) denote the position of agent \(i\), and let \(\mathbf{q}_{j}\) denote the position of landmark \(j\). 
The global landmark coverage reward is
\[
G_t=-\sum_{j=1}^{N}
\min_{i\in\{1,\ldots,N\}}
\left\lVert  \mathbf{p}_{i,t}-\mathbf{q}_{j}  \right\rVert_2.
\]

The local collision penalty of agent \(i\) is
\[
L_{i,t}=-\sum_{\substack{k=1\\k\neq i}}^{N} \mathbb{I}
\left[ \left\lVert
\mathbf{p}_{i,t}-\mathbf{p}_{k,t} \right\rVert_2 < s_i+s_k
\right],
\]
where \(s_i\) and \(s_k\) are the collision radius of the two agents. Following the environment default, we use a local-reward ratio of \(\rho=0.5\), giving
\[
r^{\mathrm{task}}_{i,t}=(1-\rho)G_t+\rho L_{i,t}.
\]

% Each agent receives an \(18\)-dimensional local observation containing its velocity and position, the relative positions of the landmarks, and the relative positions of the other agents.

Since \textsc{Spread} contains no communication channel, the total energy is
\[
    E_{i,t}=E^{\mathrm{move}}_{i,t}.
\]

\paragraph{\textbf{\textsc{Formation}.}}
The \textsc{Formation} task contains \(N=3\) agents and one central landmark. The agents must arrange themselves at equally spaced locations on a circle of radius \(r_{\mathrm{form}}=0.5\) centred on the landmark. 
Let \(\{\mathbf{q}_{1,t},\ldots,\mathbf{q}_{N,t}\}\) denote the ideal formation positions. 
Since the agents are interchangeable, they are assigned to target positions using minimum-cost bipartite matching:

\[
\pi_t^{*}
=
\arg\min_{\pi\in\Pi_N}
\sum_{i=1}^{N}
\left\lVert
\mathbf{p}_{i,t}-\mathbf{q}_{\pi(i),t}
\right\rVert_2,
\]

where \(\Pi_N\) is the set of permutations of \(N\) elements. The shared task reward is the negative mean matched distance:

\[
r^{\mathrm{task}}_{i,t}
=
-\frac{1}{N}
\sum_{i'=1}^{N}
\operatorname{clip}
\left(
\left\lVert
\mathbf{p}_{i',t}
-
\mathbf{q}_{\pi_t^{*}(i'),t}
\right\rVert_2,
0,2
\right).
\]

% The same task reward is given to every agent. Each local observation has six dimensions and contains the agent's velocity and position together with the relative position of the central landmark.
Since \textsc{Formation} contains no communication channel, the total energy is 
\[
    E_{i,t}=E^{\mathrm{move}}_{i,t}.
\]

\paragraph{\textbf{\textsc{Line}.}}
The \textsc{Line} task contains \(N=3\) agents and two landmarks. The landmarks are separated by \(1.25\) spatial units in a randomly sampled direction, and the agents must occupy \(N\) evenly spaced positions along the line segment joining them. 
Let \(\{\mathbf{q}_{1},\ldots,\mathbf{q}_{N}\}\) denote these ideal positions. The agents are again matched to target positions using
\[
\pi_t^{*}=\arg\min_{\pi\in\Pi_N}
\sum_{i=1}^{N}
\left\lVert
\mathbf{p}_{i,t}-\mathbf{q}_{\pi(i)}
\right\rVert_2.
\]

The shared task reward is
\[
r^{\mathrm{task}}_{i,t}=-\frac{1}{N}\sum_{i'=1}^{N}
\operatorname{clip}
\left(\left\lVert
\mathbf{p}_{i',t}-\mathbf{q}_{\pi_t^{*}(i')}\right\rVert_2,
0,2
\right).
\]
% Each agent receives an eight-dimensional local observation containing its velocity and position and the relative positions of the two landmarks.
Since \textsc{Line} contains no communication channel, the total energy is
\[
    E_{i,t}=E^{\mathrm{move}}_{i,t}.
\]

\paragraph{\textbf{\textsc{Reference}.}}
The \textsc{Reference} task contains two agents and three visually distinct landmarks. Each agent is assigned a target landmark, which is known only by the other agents. Both agents are simultaneous speakers and listeners.
Successful coordination therefore requires the agents to communicate their private goal information while simultaneously moving towards their respective targets.

Let \(\mathbf{q}^{*}_{i}\) denote the target landmark assigned to agent \(i\).
We use the fully cooperative setting by fixing the environment's local-reward ratio to zero. Both agents consequently receive the shared task reward

\[
r^{\mathrm{task}}_{i,t}=-\frac{1}{2}
\sum_{j=1}^{2}\left\lVert
\mathbf{p}_{j,t}-\mathbf{q}^{*}_{j}
\right\rVert_2.
\]

% Each agent receives a \(21\)-dimensional observation containing its velocity, the relative positions of the three landmarks, the colour encoding of its partner's target, and the most recent message received from the other agent.

In addition to the two-dimensional movement command \(\mathbf{f}_{i,t}\), each agent selects a binary communication gate \(\mathbf{g}_{i,t}\in\{0,1\}^{10}\) and a continuous message payload \(\mathbf{m}_{i,t}\in[0,1]^{10}\):

\[
\mathbf{a}_{i,t}=\left[
\mathbf{f}_{i,t},\,\mathbf{g}_{i,t},\,\mathbf{m}_{i,t}
\right].
\]

The transmitted message is
\[\mathbf{c}_{i,t}=\mathbf{g}_{i,t}\odot\mathbf{m}_{i,t},\]
so a message component is transmitted only when its corresponding gate is active.

Communication energy is determined by the transmitted packet length. Let \(b=8\) be the number of bits charged per active communication component. The packet length is
\[B_{i,t}=b\sum_{j=1}^{10}g_{i,t}^{j},\]

Let \(e_{\mathrm{bit}}=0.025\) be the energy cost per bit. 
The communication energy is
\[
E^{\mathrm{comm}}_{i,t}=e_{\mathrm{bit}}B_{i,t}
=0.2\sum_{j=1}^{10}g_{i,t}^{j}.
\]

The communication cost depends on the number of active components. Therefore, silence incurs no communication cost. 
The total energy objective in \textsc{Reference} is therefore

\[
r^{\mathrm{energy}}_{i,t}=
-\left(
E^{\mathrm{move}}_{i,t}+E^{\mathrm{comm}}_{i,t}
\right).
\]

\subsection{Evaluation Metrics}\label{app:eval_metrics}

We evaluate the quality of the approximate Pareto front using three standard metrics, following the formalism in~\citep{zitzler1999multiobjective, zintgraf2015quality, hayes2022practical}.

%\paragraph
{\bf Hypervolume (HV).}  
Let \(P\) be an approximate Pareto front and \(G_0\) a reference point dominated by all \(p\in P\).  The hypervolume is
\(
  \mathcal{H}(P)
  = \int_{\mathbb{R}^d} \mathbbm{1}_{H(P)}(z)\,dz,
\)
where
\(
  H(P) = \{\,z \in \mathbb{R}^n \mid \exists\,j,\;1\le j\le |P|\;:\;G_0 \preceq z \preceq P(j)\}
\). 
Here, \(P(j)\) is the \(j^{\rm th}\) solution in \(P\), the symbol \(\preceq\) denotes objective dominance, and \(\mathbbm{1}_{H(P)}\) is an indicator %Dirac 
function that equals 1 if \(z\in H(P)\) and 0 otherwise. A higher hypervolume implies a front closer to and more extensive with respect to the true Pareto front.

%\paragraph
{\bf Expected Utility (EU).}  
Let \(P\) be an approximate Pareto front and \(\Pi\) be the corresponding policy set. The expected utility metric is
\(
  \mathcal{U}(P)
  = \mathbb{E}_{\omega}\bigl[\max_{\pi\in\Pi}\;\omega^\top G^\pi_\omega\bigr].
\)
A higher EU denotes better average performance over preferences.

%\paragraph
{\bf Sparsity (SP).}  
Let \(P\) be an approximate Pareto front in a \(d\)-dimensional objective space. The sparsity metric is
\(
  S(P)
  = \frac{1}{|P|-1}
    \sum_{i=1}^d
      \sum_{k=1}^{|P|-1}
        \bigl(\tilde G_i(k) - \tilde G_i(k+1)\bigr)^2,
\)
where \(\tilde G_i\) is the sorted list of the \(i^{\rm th}\) objective values in \(P\), and \(\tilde G_i(k)\) is the \(k^{\rm th}\) entry in this sorted list. 
Lower sparsity indicates a more uniform distribution of solutions along each objective.
For a singleton solution set, SP is zero by definition but does not provide a meaningful measure of distribution. We therefore report such cases as ``--''.

\subsection{Training Details}\label{app:training_detail}

\textbf{Training budget and preference handling.}
All methods are trained using a fixed budget of \(5\times10^{6}\) joint environment steps for each environment and random seed. We use 3 random seeds for each environment, an episode horizon of \(H=25\), and a discount factor of \(\gamma=0.99\). Except for \textit{MO-MIX}, all methods use linear scalarisation and are trained separately for \(K=21\) uniformly spaced team preferences with an interval of \(0.05\), with the total training budget divided evenly across these preferences. For the PPO-based baselines, the vector-valued reward is scalarised using the fixed team preference. In our methods, the scalarisation weights \(\boldsymbol{\omega}_{\mathrm{team}}\) represent the user-specified team preference and define the desired system-level objective trade-off, while the learned individual preferences are used within the agent learning process. \textit{MO-MIX}, in contrast, uses a single preference-conditioned network trained with the complete \(5\times10^{6}\)-step budget, with the requested team preference provided directly as an input and training preferences sampled using the same interval of \(0.05\).

\textbf{PPO-based baselines.}
We consider three PPO-based baselines with progressively richer information available during training.
\textsc{DTDE-PPO} uses fully decentralised training and execution, with each agent learning independently from its local observation and scalarised reward.
\textsc{CTDE-PPO-V} introduces a centralised state-value critic during training, while \textsc{CTDE-PPO-Q} instead uses a centralised joint-action critic.
All three methods retain decentralised actors during execution and use the same training budget and preference settings described above.
Their main hyperparameters are summarised in Table~\ref{tab:ppo_hyperparameters}.

\begin{table}[t]
\centering
\caption{Hyperparameters used by the PPO-based baselines.}
\label{tab:ppo_hyperparameters}
\scalebox{0.8}{
\begin{tabular}{lccc}
\toprule
\textbf{Hyperparameter}
& \textbf{\textsc{DTDE-PPO}}
& \textbf{\textsc{CTDE-PPO-V}}
& \textbf{\textsc{CTDE-PPO-Q}} \\
\midrule
Actor learning rate                     & \(3\times10^{-4}\) & \(3\times10^{-4}\) & \(3\times10^{-4}\) \\
Critic learning rate                    & \(3\times10^{-4}\) & \(3\times10^{-4}\) & \(3\times10^{-4}\) \\
PPO clipping coefficient                & \(0.2\) & \(0.2\) & \(0.2\) \\
Entropy coefficient                     & \(0.01\) & \(0.01\) & \(0.01\) \\
Optimisation epochs                     & \(4\) & \(4\) & \(4\) \\
Minibatch size                          & \(256\) & \(256\) & \(256\) \\
Rollout horizon                         & \(512\) & \(512\) & \(512\) \\
Actor hidden layers                     & \(128,128\) & \(128,128\) & \(128,128\) \\
Critic hidden layers                    & \(256,256\) & \(256,256\) & \(256,256\) \\
\(\lambda\)                         & \(0.8\) & \(0.8\) & \(0.8\) \\

\bottomrule
\end{tabular}
}
\end{table}

\textbf{MO-COMA, \algoname-S, and \algoname-D.}
\textit{MO-COMA} extends COMA~\citep{foerster2018counterfactual} to vector-valued rewards and serves as the base implementation for both proposed \algoname{} variants. 
For continuous action spaces, the counterfactual expectation is approximated using 8 actions sampled from the current policy. 
\textit{\algoname-S} and \textit{\algoname-D} are built directly on the same \textit{MO-COMA} implementation and introduce the static and dynamic meta-gradient preference-adaptation mechanisms described in the main text, respectively.
The shared implementation settings and additional meta-learning hyperparameters are summarised in Table~\ref{tab:coma_hyperparameters}.

\begin{table}[t]
\centering
\caption{Hyperparameters used by \textsc{MO-COMA}, \textsc{MIRL-S}, and \textsc{MIRL-D}.}
\label{tab:coma_hyperparameters}
\scalebox{0.8}{
\begin{tabular}{lccc}
\toprule
\textbf{Hyperparameter}
& \textbf{\textsc{MO-COMA}}
& \textbf{\textsc{MIRL-S}}
& \textbf{\textsc{MIRL-D}} \\
\midrule
Actor learning rate                     & \(5\times10^{-4}\) & \(5\times10^{-4}\) & \(5\times10^{-4}\) \\
Critic learning rate                    & \(5\times10^{-4}\) & \(5\times10^{-4}\) & \(5\times10^{-4}\) \\
Episode batch size                      & \(8\) & \(8\) & \(8\) \\
\(\lambda\)                         & \(0.8\) & \(0.8\) & \(0.8\) \\
Actor recurrent hidden size             & \(64\) & \(64\) & \(64\) \\
Critic hidden size                      & \(128\) & \(128\) & \(128\) \\
Counterfactual action samples           & \(8\) & \(8\) & \(8\) \\
Meta learning rate                      & -- & \(10^{-4}\) & \(10^{-4}\) \\
Meta-trace decay $\mu$                       & -- & \(0\) & \(0\) \\
Dynamic preference hidden size          & -- & -- & \(64\) \\
\bottomrule
\end{tabular}
}
\end{table}

\textbf{MO-MIX.}
We implement \textit{MO-MIX} following~\citep{hu2023momix} and use the parameters described in the original paper.
As \textit{MO-MIX} operates on discrete action spaces, we discretise the continuous movement actions into five choices: stay, left, right, down, and up.
For \textit{Reference}, these movement actions are combined with 11 communication choices, resulting in 55 discrete actions per agent.
The model is trained as a single preference-conditioned policy using the preference sampling scheme described above.
To improve training stability, we normalise the positive mixing-network weights while preserving the monotonicity constraint.
The main hyperparameters are summarised in Table~\ref{tab:momix_hyperparameters}.

\begin{table}[!t]
\centering
\caption{Hyperparameters used by \textsc{MO-MIX}.}
\label{tab:momix_hyperparameters}
\scalebox{0.85}{
\begin{tabular}{lc}
\toprule
\textbf{Hyperparameter} & \textbf{Value} \\
\midrule
Learning rate                    & \(5\times10^{-6}\) \\
Episode batch size               & \(64\) \\
Replay capacity                  & \(12000\) episodes \\
Agent recurrent hidden size      & \(64\) \\
Mixing hidden size               & \(32\) \\
Training preference interval     & \(0.05\) \\
Evaluation preference interval   & \(0.0125\) \\
\bottomrule
\end{tabular}
}
\end{table}

\subsection{Computational Resources}
All experiments were conducted on a local workstation equipped with an NVIDIA GeForce RTX 5090 GPU (32 GB VRAM), an Intel Core Ultra 7 270K CPU, and 64 GB of system memory, running Ubuntu 24.04 LTS. Each experimental run used a single GPU.

\section{Additional Results}\label{app:additional_exp}
In this section, we provide additional results to further analyse how meta-gradient-informed preference adaptation operates during training and execution.
For the two-objective setting, we write 
$\boldsymbol{\omega}_{\mathrm{team}}=(\omega_{\mathrm{team,task}}, \omega_{\mathrm{team,energy}})$for the team preference and
$\boldsymbol{\omega}_i=(\omega_{i,\mathrm{task}},\omega_{i,\mathrm{energy}})$ for the individual preference of agent $i$.
Since the two components sum to one, we report only the first task component in this section.

We first examine how individual preferences are learned during
training, then compare the static and dynamic variants, and finally
inspect the context-dependent preferences generated by
\algoname{}-D during execution.

\begin{figure}[ht]
    \centering
    \begin{subfigure}[t]{0.23\textwidth}
        \centering
        \includegraphics[width=\linewidth]
        {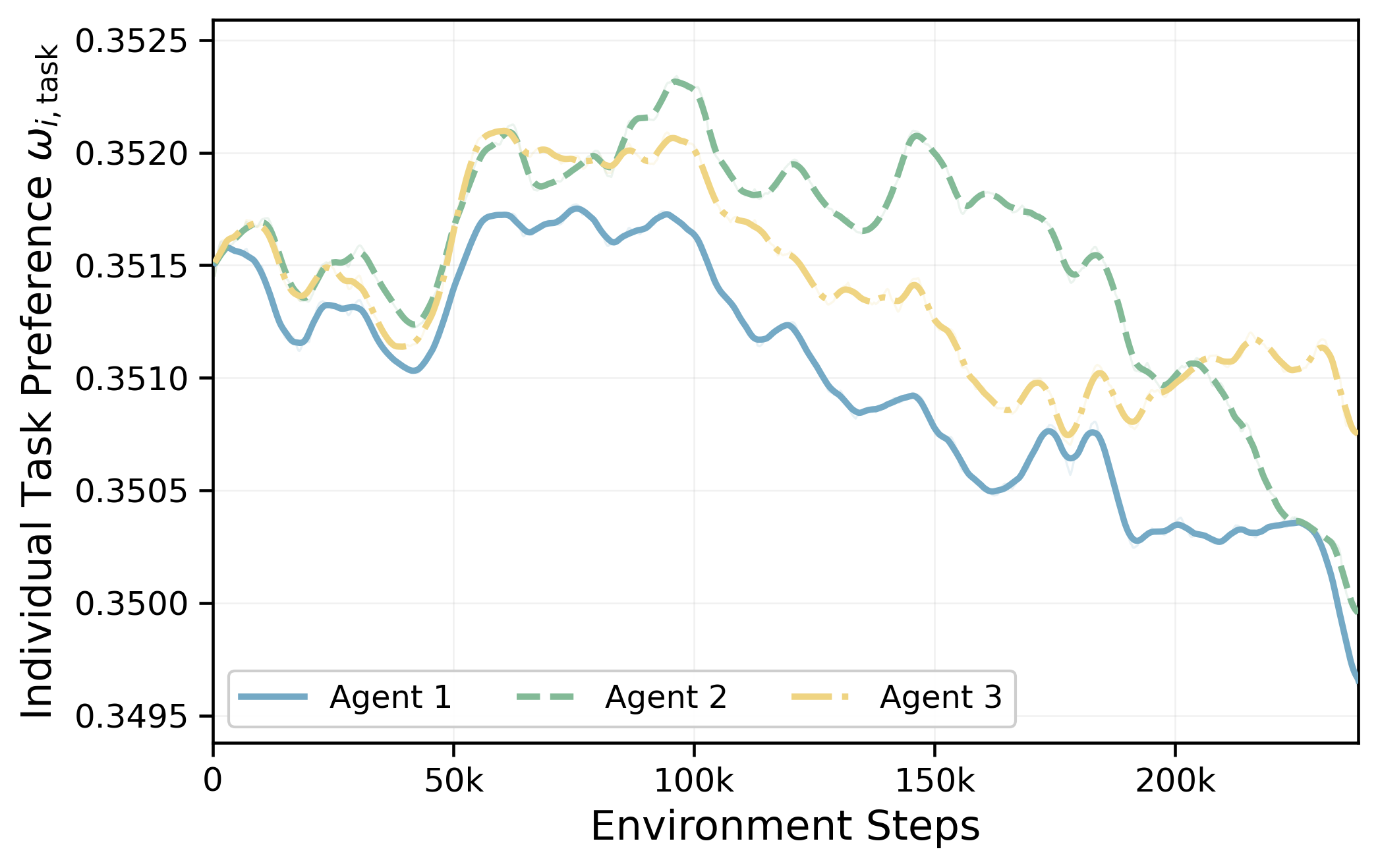}
        \caption{
        $\omega_{\mathrm{team,task}}=0.35$
        }
        \label{fig:mirl_s_meta_w035}
    \end{subfigure}
    \hfill
    \begin{subfigure}[t]{0.23\textwidth}
        \centering
        \includegraphics[width=\linewidth]
        {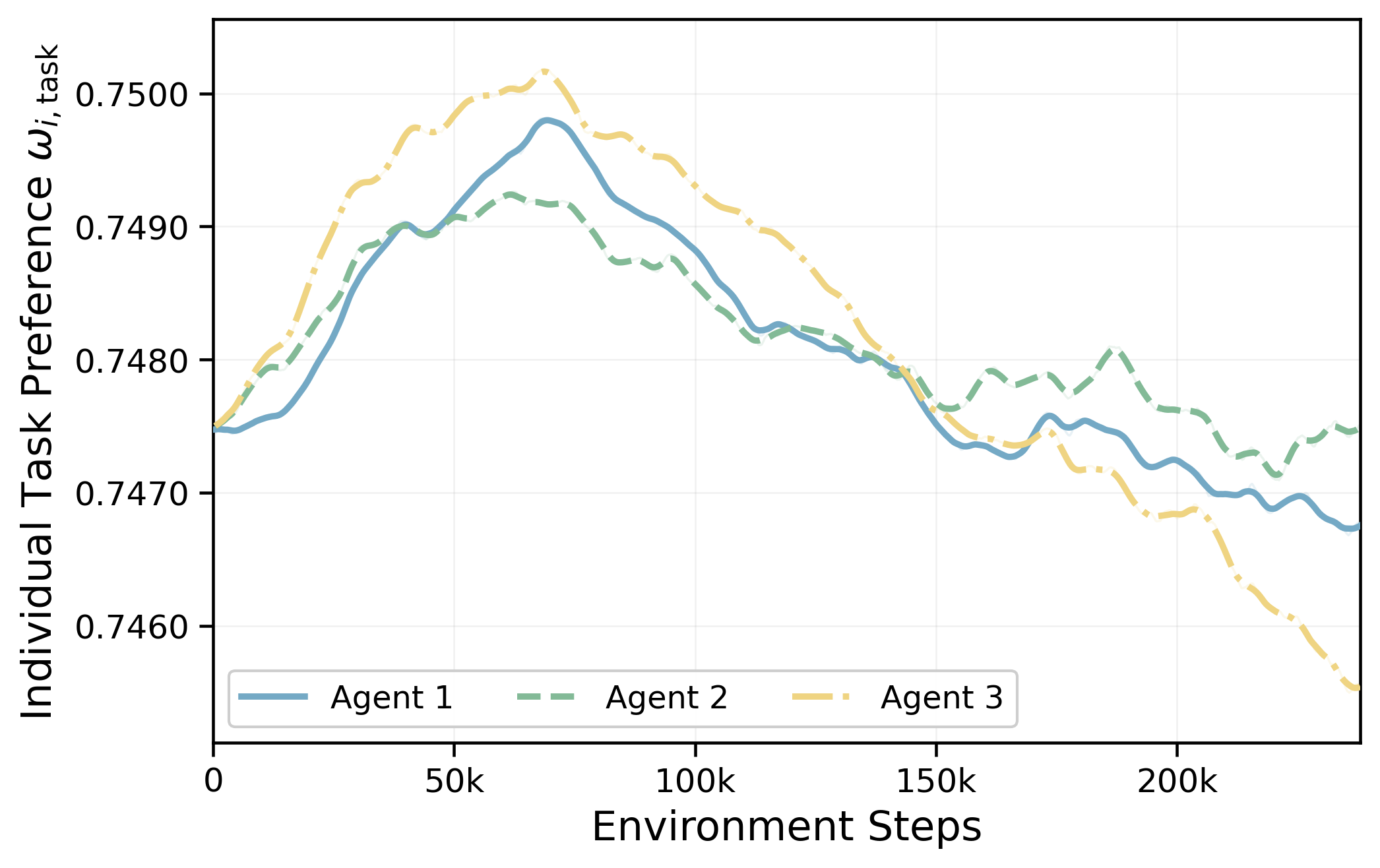}
        \caption{
        $\omega_{\mathrm{team,task}}=0.75$
        }
        \label{fig:mirl_s_meta_w075}
    \end{subfigure}
    \caption{Individual task preferences learned by \algoname{}-S during training on \textit{Line}. }
    \label{fig:mirl_s_meta_preferences}
\end{figure}

% We first examine how the learned individual preferences change during training.
Figure~\ref{fig:mirl_s_meta_preferences} shows two representative runs of \algoname{}-S on \textit{Line} under different team preferences. 
The individual preferences are initialised close to the corresponding team preference, but gradually change in different directions as training proceeds.
In particular, the preferences of different agents become separated, showing that the meta-gradient updates learn agent-specific objective emphases rather than preserving the common initial preference.

\begin{figure}[h]
    \centering
    \begin{subfigure}[t]{0.23\textwidth}
        \centering
        \includegraphics[width=\linewidth]
        {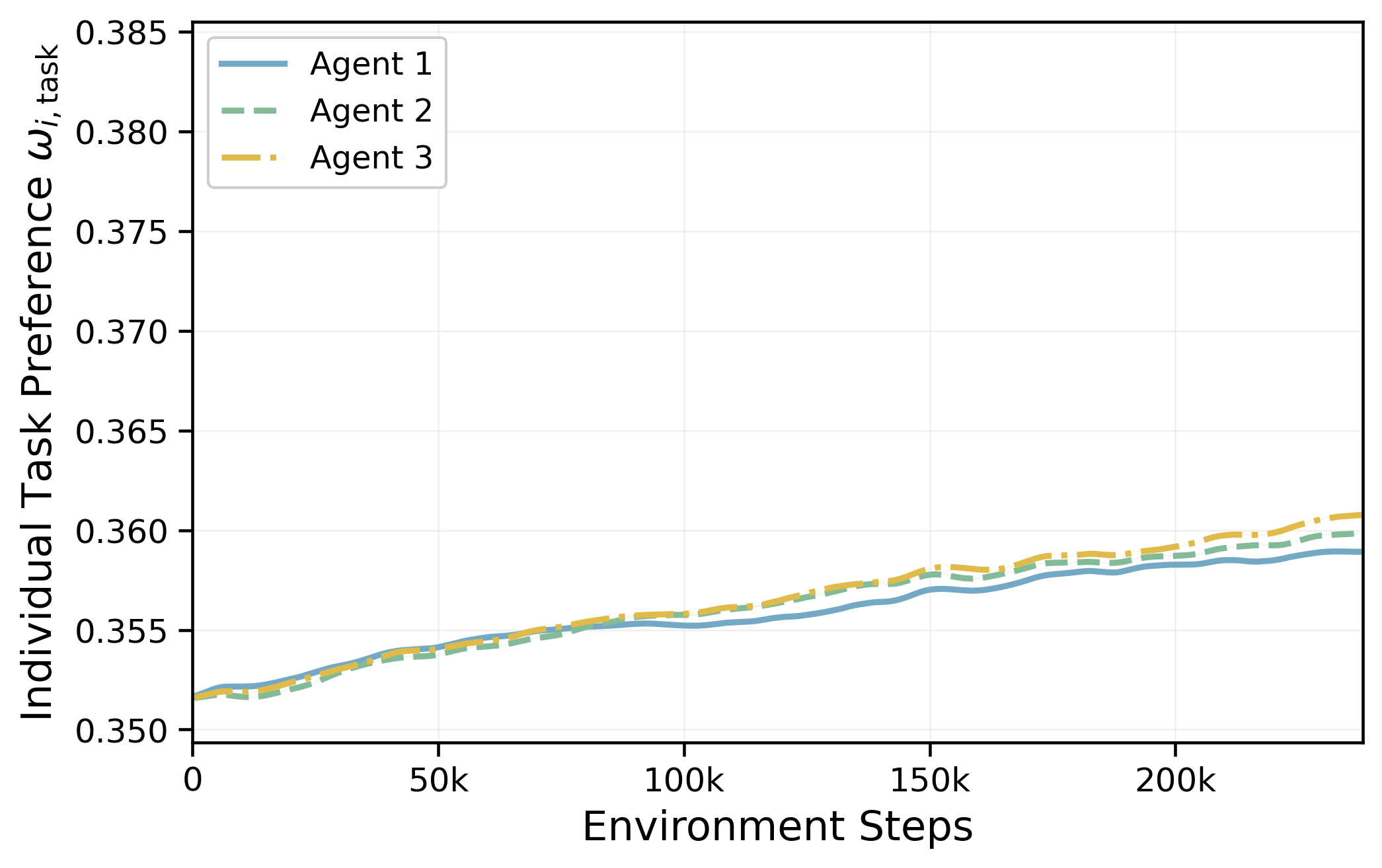}
        \caption{\algoname{}-S}
        \label{fig:mirl_s_spread_meta}
    \end{subfigure}
    \hfill
    \begin{subfigure}[t]{0.23\textwidth}
        \centering
        \includegraphics[width=\linewidth]
        {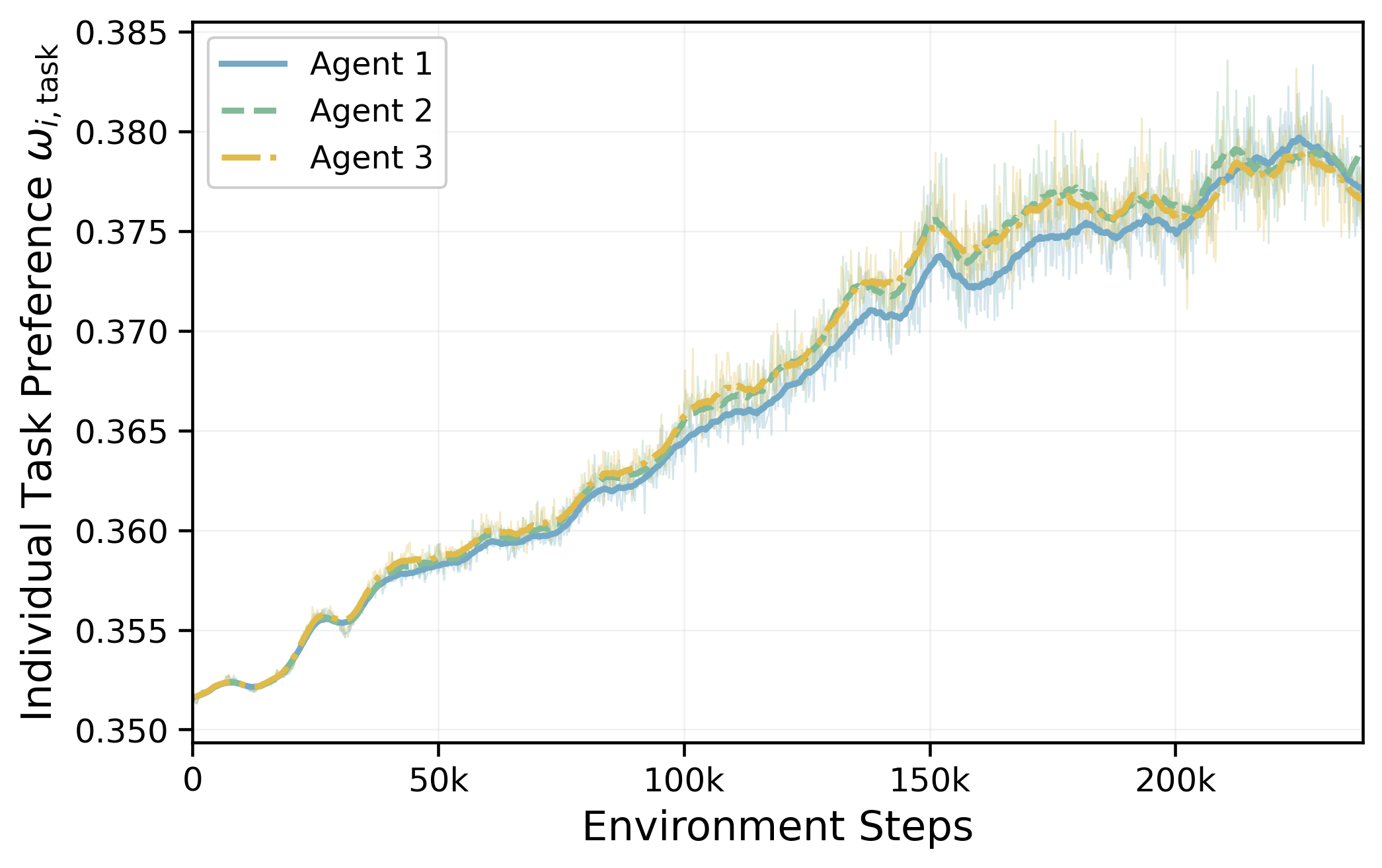}
        \caption{\algoname{}-D}
        \label{fig:mirl_d_spread_meta}
    \end{subfigure}
    \caption{
    Individual task preferences learned by \algoname{}-S and \algoname{}-D on \textit{Spread}, with     $\omega_{\mathrm{team,task}}=0.35$.
    }
    \label{fig:mirl_sd_meta_comparison}
\end{figure}

Having shown that \algoname{}-S learns agent-specific preferences, we next compare the preference adaptation of \algoname{}-S and \algoname{}-D under the same training setting.
Figure~\ref{fig:mirl_sd_meta_comparison} shows a representative run on \textit{Spread} with  $\boldsymbol{\omega}_{\mathrm{team}}=(0.35,0.65)$.
For \algoname{}-S, each curve represents the individual task preference learned for one agent.
For \algoname{}-D, the preference is generated dynamically at each timestep, so we report the mean task preference produced for each agent over the corresponding training batch.

Both variants progressively adapt their individual preferences away from the common initialisation.
In \algoname{}-S, the learned agent preferences gradually separate during training.
In \algoname{}-D, the generated preferences show a larger overall shift and greater short-term variation, consistent with the additional flexibility of the context-dependent preference controller.
These results confirm that the meta-gradient updates actively adapt both the static preference parameters in \algoname{}-S and the dynamic preference controller in \algoname{}-D.

\begin{figure}[h]
    \centering
    \begin{subfigure}[t]{0.23\textwidth}
        \centering
        \includegraphics[width=\linewidth]
        {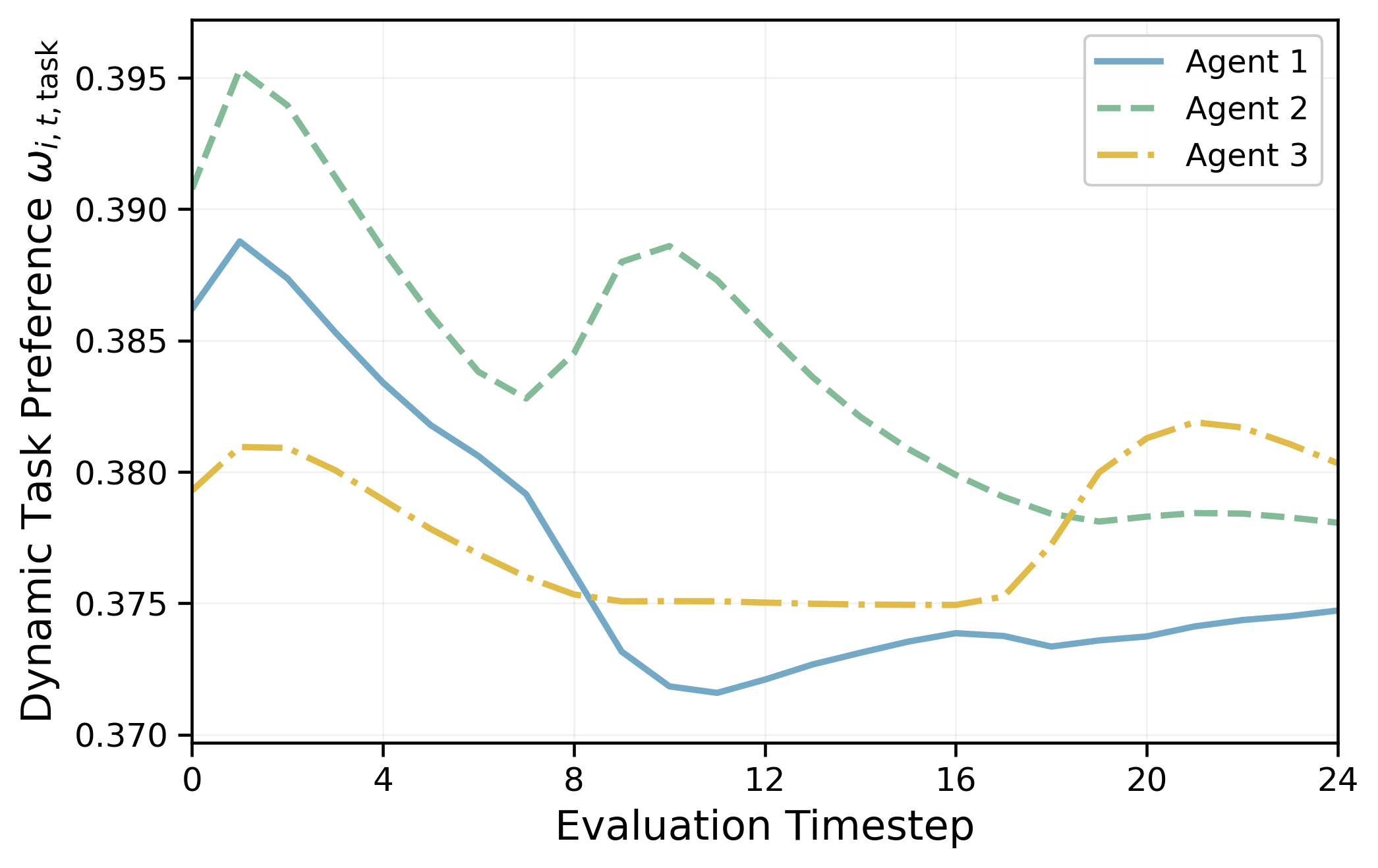}
        \caption{Evaluation episode 1}
        \label{fig:mirl_d_execution_ep1}
    \end{subfigure}
    \hfill
    \begin{subfigure}[t]{0.23\textwidth}
        \centering
        \includegraphics[width=\linewidth]
        {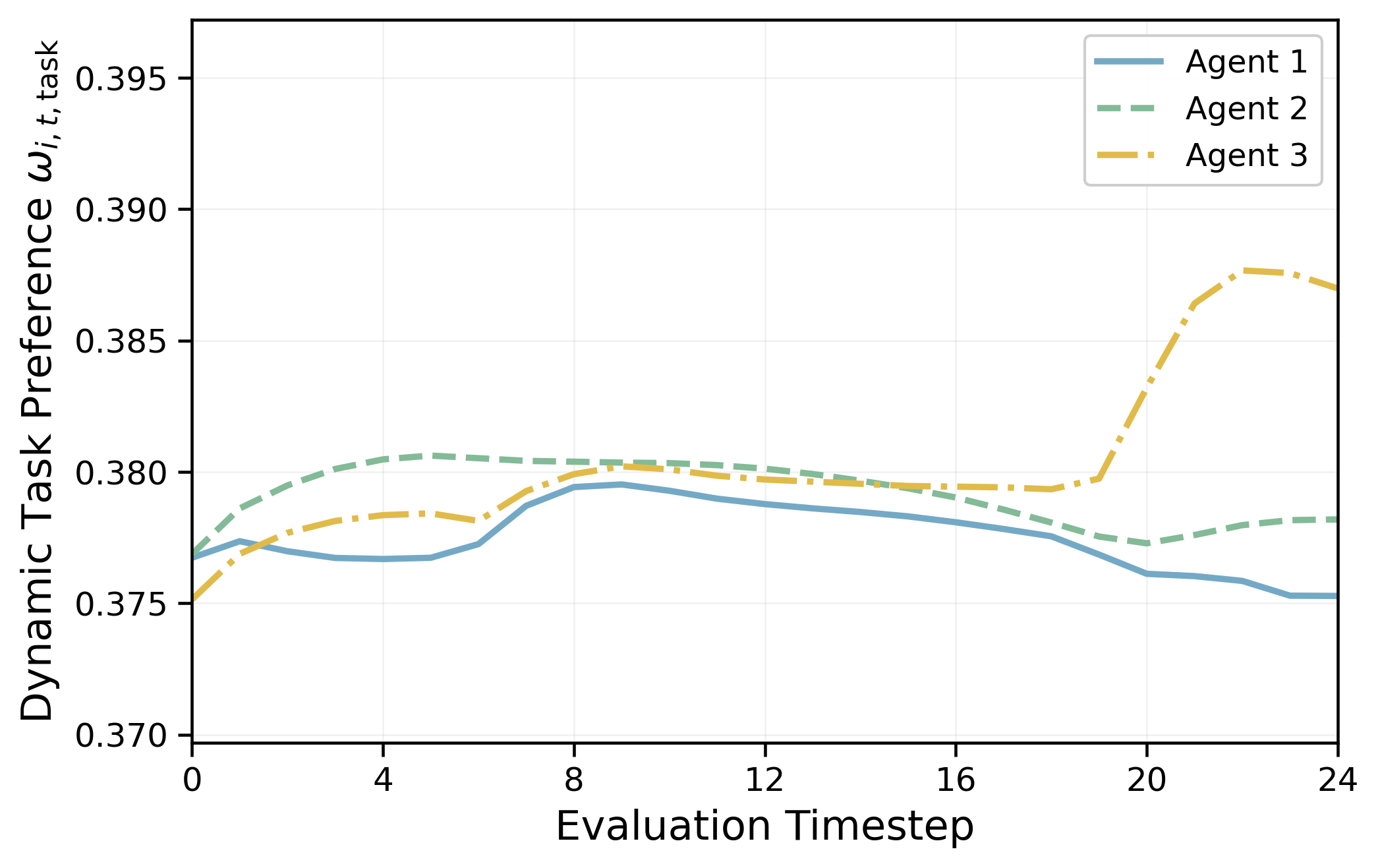}
        \caption{Evaluation episode 2}
        \label{fig:mirl_d_execution_ep2}
    \end{subfigure}
    \caption{
    Individual task preferences generated by \algoname{}-D during two example evaluation episodes on \textit{Spread} under the fixed team preference $\omega_{\mathrm{team,task}}=0.35$.
    }
    \label{fig:mirl_d_execution_preferences}
\end{figure}

While the previous analysis focuses on preference adaptation during training, \algoname{}-D is designed to generate context-dependent preferences during execution.
We therefore further examine how its individual preferences vary within an evaluation episode.

Unlike \algoname{}-S, which maintains one learned preference for each agent, \algoname{}-D generates an individual preference $\boldsymbol{\omega}_{i,t}$ at each timestep from the agent's recurrent hidden state.
Figure~\ref{fig:mirl_d_execution_preferences} shows two example evaluation episodes from the same trained policy on \textit{Spread}, with the team preference fixed to $\boldsymbol{\omega}_{\mathrm{team}}=(0.35,0.65)$.
Since the two preference components sum to one, we again report only the task component $\omega_{i,t,\mathrm{task}}$.
The generated preferences vary across both agents and timesteps, while the team preference remains fixed throughout each episode.
Moreover, the two episodes produce different preference trajectories, showing that \algoname{}-D generates context-dependent individual preferences rather than assigning each agent a fixed preference.
Since no parameter updates are performed during evaluation, these changes reflect the dynamic context-dependent preference generation of the trained controller during execution.

\section{Limitations}
The learned individual preferences provide a potentially interpretable description of how different agents emphasise the underlying objectives, and our current results show that these preferences adapt differently across agents and cooperative contexts.
However, the present benchmark tasks are not designed to cleanly separate distinct behavioural tendencies or cooperative roles, which limits our ability to quantitatively relate changes in the learned preferences to specific changes in agent behaviour.
A more systematic analysis would therefore require environments in which such behavioural tendencies can be independently identified and measured.
This represents an interesting direction for future work, particularly for studying whether the learned preferences can provide a meaningful and interpretable characterisation of agent roles and contributions.

Our current evaluation is also limited to two-objective cooperative tasks with relatively small agent populations.
Extending the analysis to larger multi-agent systems and higher-dimensional objective spaces would provide a broader assessment of the scalability and generality of the proposed approach.

%%%%%%%%%%%%%%%%%%%%%%%%%%%%%%%%%%%%%%%%%%%%%%%%%%%%%%%%%%%%%%%%%%%%%%%%

\end{document}